\documentclass[12pt]{article}
\usepackage{amsfonts,amsbsy,latexsym,amssymb,amscd,amstext}

\newcommand{\be}{\begin{equation}}
\newcommand{\ee}{\end{equation}}

\newcommand{\Fmunu}{{\bf F}_{\mu \nu}(x)}
\newcommand{\Ftmunu}{\widetilde{\bf F}_{\mu \nu}(x)}

\newcommand{\SUT}{SU(2)}

\newcommand{\Omegamu}{\Omega_{\mu}} 
\newcommand{\Omeganu}{\Omega_{\nu}}

\newcommand{\nmunu}{n_{\mu \nu}}

\newcommand{\half}{\frac{1}{2}}
\newcommand{\etabarmunu}{\overline{\eta}_{\mu\nu}}

\newcommand{\bea}{\begin{eqnarray}}
\newcommand{\eea}{\end{eqnarray}}
\input{epsf.tex}

\newcommand{\Real}{R}

\begin{document}

\vskip -4cm

\begin{flushright}
FTUAM-00-16

IFT-UAM/CSIC-00-23
\end{flushright}

\vskip 0.2cm

{\Large
\centerline{\bf Perturbative construction of self-dual configurations on 
the torus }}
%\centerline{{\bf }}
\vskip 0.3cm

\centerline{\qquad \ \ 
M. Garc\'{\i}a P\'erez{$^\dag$,  A. Gonz\'alez-Arroyo{$^\dag$}{$^\ddag$} and C. Pena{$^\dag$} \\ }}
\vskip 0.3cm

\centerline{{$^\dag$}Departamento de F\'{\i}sica Te\'orica C-XI,}
\centerline{Universidad Aut\'onoma de Madrid,}
\centerline{Cantoblanco, Madrid 28049, SPAIN.}
\vskip 10pt
\centerline{{$^\ddag$}Instituto de F\'{\i}sica Te\'orica C-XVI,}
\centerline{Universidad Aut\'onoma de Madrid,}
\centerline{Cantoblanco, Madrid 28049, SPAIN.}

\vskip 0.8cm

\begin{center}
{\bf ABSTRACT}
\end{center}
We develop a perturbative expansion which allows the construction of 
non-abelian self-dual SU(2) Yang-Mills field 
configurations on the four-dimensional torus with topological charge $1/2$.
The expansion is performed around the constant field strength abelian
solutions found by 't Hooft. 
Next to leading order  calculations are compared with numerical 
results obtained with lattice gauge theory techniques.

%\vskip 1.0 cm
%\begin{flushleft}
%PACS: 

%Keywords:  Instantons, .
%\end{flushleft}

\newpage

\section{Introduction}

Self-dual Yang-Mills fields are fascinating mathematical objects
that play an important role in both Physics and Mathematics. In the
Physics literature they emerged through the introduction of the 
BPST instanton~\cite{BPST},
the minimum action 
configuration in the sector of topological charge $Q$ equal to one. This
triggered the joint effort of physicists and mathematicians in the search 
for  multiinstanton configurations. This had its reward with the 
ADHM formalism~\cite{ADHM}, a general set up for the construction of
self-dual configurations with vanishing field strength at infinity
(which, from the topological point of view, corresponds to fields
living on a four-sphere). However,
there are certain instances in which other types of boundary conditions are relevant. 
For example, in considering  finite temperature field theory, one is interested in 
studying configurations which are periodic in thermal time. This periodicity 
might also be used as a device to study monopole-like objects.  In the same
fashion, additional periodicities might have other uses and interpretations,
as argued in Ref.~\cite{vortex}. This justifies an initial interest in the study of self-dual 
configurations which are periodic in all euclidean space-time directions. 
Geometrically, it  corresponds to the study of self-dual gauge fields on the 
torus. It might be considered as   the next step after the case of
gauge fields on the sphere. 
 The study of gauge 
fields on the torus (see~\cite{Review} for a review of this topic) brings in a new topological richness,
whose appearance, physical interpretation and usefulness was put forward by
't Hooft~\cite{Tho1}. In addition, periodic self-dual configurations constitute a 
simple mathematical example of  dense multiinstanton 
configurations, which, as advocated in Ref.~\cite{TPA},
 might turn out to be a better description of the 
confining vacuum than other dilute multiinstanton pictures. 

The construction of self-dual configurations on the torus has met very limited
success. For particular values of the torus sizes, there is a class of
solutions which is known since  early times~\cite{Tho2}. These configurations
have constant field strength, and they are, in some sense, of an abelian nature:
all the spatial components of the electric and magnetic fields are parallel
in colour space. For other torus sizes, we know, through numerical
techniques~\cite{Numerical,overimproved,vortex}, that the solutions are quite different. They turn
out to be lumpy and very
non-abelian: different electric field components are mutually orthogonal in
colour space at the center of the lump. Fortunately,
there has recently been substantial progress in studying the mixed situation
of instanton configurations which are periodic in only a few directions.
In particular, in the case of only one compactified direction, corresponding
to the physical situation of finite temperature, the most general solution of
topological charge one, the caloron, has been found \cite{vanbaals1}.
Considerable progress has also been achieved for the case of gauge fields
on $T^2\times R^2$ (doubly periodic instantons), by the work of mathematicians~\cite{jardim}
and physicists~\cite{vortex,vortexn,wipf}. In these developments
 a crucial role is played by
the Nahm transform, a duality transformation which maps
self-dual configurations on tori with dual sizes \cite{Nahm,PBPvB}. 
For  the $T^3\times R$ case, the calculation of the abelian Nahm dual of the 
charge one instanton~\cite{vanbaalt3} represents a step forward. The
caloron, the doubly periodic instantons and  the $T^3\times R$ instanton 
can be viewed 
as limiting cases of configurations on the torus where either three, two  or
one directions are taken to be very large with respect to the others.

This paper is a step towards an analytical understanding of non-constant 
field strength self-dual configurations on the torus. The strategy is to 
consider torus sizes   which depart only slightly from those in which 
there exists a  self-dual constant field strength solution. Our
construction is based upon perturbing around the constant solution      
(at a torus size in which it is not self-dual), and
imposing self-duality to the resulting configuration. A systematic 
perturbative expansion arises that allows the construction of the self-dual 
solution. We show that  the solution exists order by order and is unique
up to gauge transformations and space-time translations. This is done in section 2. 
Our approach is intimately related to the study of van Baal~\cite{perturb1},
who considered perturbations around constant field strength solutions.

In section 3 we proceed to compare the lowest non-trivial order results 
obtained from our expansion with the exact solution obtained by numerical 
methods. This serves to quantify the rate of convergence of the series for
various  torus sizes.  Finally, in section 4, we  investigate
the interplay of our perturbative expansion with the Nahm transform. 
The paper is closed by section 5, where the conclusions and possible
extensions are presented.

\section{The construction}
Let us consider $\SUT$ gauge fields living on a torus of size
 $l_0\times l_1\times l_2\times l_3$. Under a translation by
 one of the periods, the gauge potentials and fields transform by  a gauge
 transformation:
 \be
 A_{\nu}(x+ e_{\mu})= [\Omegamu(x)]  A_{\nu}(x)\quad ,
 \ee
 where $e_{\mu}$ is a 4-vector of length $l_{\mu}$  along  the $\mu$-th direction, and
 $\Omegamu$ are the twist matrices. The compatibility conditions 
 of the previous equations are:
 \be
 \Omegamu(x+ e_{\nu})\, \Omeganu(x)=\exp\{  \pi \imath \nmunu \}\,
\Omeganu(x+ e_{\mu})\, \Omegamu(x)\quad ,
 \ee
 where the  elements of the antisymmetric twist tensor $\nmunu$ are
  integers defined modulo 2. In what follows we will choose
  $n_{0 3}=n_{1 2 }=-n_{3 0}=-n_{2 1 }=1$, with the remaining     
 components being zero. For the twist matrices we will take
 \be
 \Omegamu(x)=\exp\{\imath\frac{\pi}{2}\nmunu \frac{x_{\nu}}{l_{\nu}} \tau_3
\}\quad ,
 \ee
which is consistent with our choice of twist tensor. The symbols $\tau_i  $ label the  Pauli matrices.      

For  torus sizes such that $l_0  l_3= l_1 l_2$, there exist 
self-dual configurations satisfying the previous boundary conditions and 
having constant field strength. 
What we will do is to consider a slight deviation from this situation
controlled by the
parameter:
\be
\label{Deltadef}
\Delta=\frac{l_0 l_3-l_1 l_2}{\sqrt{V}}\quad ,
\ee
where $V= l_0  l_1 l_2  l_3$ is the torus volume. 
Without loss of generality we can assume that $\Delta$ is positive.
In this case there exists  a constant field strength
configuration with vector potential:
\be
\label{Bdef}
B_{\mu}(x)= -\frac{\pi}{2}\nmunu \frac{x_{\nu}}{l_{\mu} l_{\nu}} \tau_3\quad .    
\ee
This gauge potential gives rise to a field strength of the form $G_{\mu\nu}\tau_3$, where:
\be
G_{\mu \nu}= \pi \frac{\nmunu}{l_{\mu} l_{\nu}} \quad ,
\ee
The only non-zero components are $G_{0 3}$ and $G_{1 2 }$, which become of
equal magnitude at $\Delta=0$, rendering the solution self-dual.
With our choice of twist the constant field strength configuration has
topological charge $Q=1/2$ and, for $\Delta=0$,
total action $4\pi^2$.

Now, let us consider perturbing around this  gauge potential:  
\be
\label{Adef}
A_{\mu}(x)=B_{\mu}(x)+S_{\mu}(x)\,\tau_3   + W_{\mu}(x)\,   \tau_{+}
+W_{\mu}^{*}(x)\, 
  \tau_{-}\quad ,
\ee
where we have decomposed the additional field into different colour components.
 The matrices $  \tau_{\pm}=\half(  \tau_1 \pm \imath   \tau_2)$ are standard. 
 The boundary conditions on
the gauge fields translate into the real function $S_{\mu}(x)$ being  periodic  on
the box, and the complex function $W_{\mu}(x)$ satisfying
\be
W_{\rho}(x+ e_{\mu})=\exp\{\imath \pi\, \nmunu \frac{x_{\nu}}{l_{\nu}}\}\
W_{\rho}(x)\quad .
\ee
We will make the following  gauge choice, consistent  with  
these boundary conditions (the background field gauge):
\be
\label{gfcond}
 \partial_{\mu}\, A_{\mu}(x) -\imath [ B_{\mu}(x), A_{\mu}(x) ] =0\quad .
\ee

We will now demand that the resulting gauge field is self-dual
\bea
\label{selfdual}
&&\Fmunu-\Ftmunu=0,\\
\nonumber
&&\Ftmunu = \half \epsilon_{\mu\nu\rho\sigma} 
{\bf F}_{\rho \sigma}(x) \quad, \quad {\rm with} \quad \epsilon_{0123}=1 \quad, 
\eea
which will be interpreted as equations for the functions $S_{\mu}(x)$ and
$W_{\mu}(x)$. 
The best way to express these equations, together with the gauge fixing
condition, is to use the matrices $\sigma_{\mu}\equiv({\mathbf
{I}},-\imath \vec{  \tau})$ and $\overline{\sigma}_{\mu}\equiv({\mathbf
I},\imath \vec{  \tau})=\sigma_{\mu}^{\dagger}$. These matrices 
satisfy:
\be
\overline{\sigma}_{\mu} \sigma_{\nu} = \etabarmunu^{\alpha}
\sigma_{\alpha}\quad ,
\ee
where $\etabarmunu^{\alpha}$ is the `t Hooft symbol,  such that $\overline{\eta}^{0}_{\mu \nu}=
\delta_{\mu \nu}$ and the  $\overline{\eta}^{i}_{\mu \nu}$ are a basis of the
antiself-dual tensors. Now contracting $\Fmunu$ with
$\overline{\sigma}_{\mu} \sigma_{\nu}$, we project out the
self-dual part. Hence,  we might rewrite  equation (\ref{selfdual}) as follows:
\bea
\label{Seqn}
&\overline{\partial}S=-\frac{\lambda}{2}\widehat{G}+\frac{\imath}{2} (W_c^{\dagger}W_c -W^{\dagger}W)\\
\label{Weqn}
&\overline{D}W=\imath ( S^{\dagger}W - W_c^{\dagger}S)\quad ,
\eea
where $S=S_{\mu}(x)\sigma_{\mu}$ and
$W=W_{\mu}(x)\sigma_{\mu}$ are $2\times 2$ matrices,
$S^{\dagger}$, $W^{\dagger}$ their adjoints and the parameter $\lambda$ is
equal to $1$. The matrix $W_c$ is the charge
conjugate:
\be
W_c=  \tau_2 W^{*}   \tau_2\quad .
\ee
The matrix $\widehat{G}\equiv  G_{\mu \nu} \overline{\sigma}_{\mu}
\sigma_{\nu}$ is given by:
\be
\widehat{G}= 2 \pi \imath \frac{ \Delta}{\sqrt{V}}  \tau_3 \quad . 
\ee
It vanishes when $\Delta=0$, since then the constant field strength 
configuration is self-dual. 
Finally, we define the following differential operators:
\bea
\label{defpart}
\partial &=& \sigma_{\mu}\partial_{\mu} \ , \\
\label{defbpart}
\overline{\partial} &=& \overline{\sigma}_{\mu}\partial_{\mu} \ , \\
\label{defD}
D &=& \sigma_{\mu}D_{\mu} \ , \\
\label{defbD}
\overline{D} &=& \overline{\sigma}_{\mu}D_{\mu} \ , \\
\label{defDmu}
D_{\mu} &=& \partial_{\mu} + \imath \pi\frac{x_{\nu}\nmunu}{l_{\mu}l_{\nu}} \ , 
\eea

Since at  $\Delta=0$ the correction terms $W$ and $S$ vanish, we can think
of  $\Delta$ as a perturbation parameter. Rigorously speaking this is not
quite so, because $\Delta$ depends on the torus sizes, and they enter also in the
boundary conditions. To keep a truly perturbative parameter in our expansion
we introduced in Eq.~(\ref{Seqn}) the parameter $\lambda$, whose
interpretation will be clear later. Our goal will be to solve
Eqs.~(\ref{Seqn})-(\ref{Weqn}) for arbitrary values of $\lambda$, as a
perturbative expansion in powers of $\lambda \Delta$. For that  
we have to  expand the 
unknown matrices $S$,$W$ in powers of $ \sqrt{\lambda \Delta}$. We see that 
the equations are 
consistent with $W$ carrying odd powers and $S$ even powers:
\bea
\label{Wexp}
W &=& \frac{1}{\sqrt{\lambda\,\Delta}} \sum_{k=1}^{\infty} (\lambda\,\Delta)^k W^{(k)} \\
\label{Sexp}
S &=&  \sum_{k=1}^{\infty}(\lambda\, \Delta)^k S^{(k)} \quad .
\eea
In the following paragraphs we will show that it is possible to solve 
the set of equations (\ref{Seqn})-(\ref{Weqn}) order by order in 
$\sqrt{\lambda \Delta}$. 
Finally, setting $\lambda=1$ one recovers the solution of the
self-duality condition. However, even for this value higher orders in
the expansion are suppressed by powers of $\sqrt{\Delta}$. Thus, as we will
verify later,  we expect the first few terms of the expansion to approximate 
the self-dual solutions for small values of $\Delta$.

On the other hand, the  solution for arbitrary $\lambda$ can be interpreted 
as the solution of the following
modified self-duality equation:
\be
\Fmunu-\Ftmunu=(1-\lambda) (G_{\mu \nu}-\widetilde{G}_{\mu \nu}) \tau_3\quad . 
\ee
For $\lambda=0$ the constant field strength configuration is a solution,
while for $\lambda=1$ we recover the self-duality equations.

Now let us address solving the equations order by order in $\lambda$. 
Notice first that $W^{(1)}$ satisfies the equation:
\be
\label{W0eqn}
\overline{D} W^{(1)} =0\quad .
\ee
This equation has non-zero regular solutions, as can be deduced from the index
theorem. As we will see later the general solution has the form:
\be
\label{W0sol}
W^{(1)}= \Psi(x)\, \left( \begin{array}{c c} K^{(1)}\ & Q^{(1)} \\ 0 \ & 0
\end{array} \right) \quad ,
\ee
where $K^{(1)}$ and $Q^{(1)}$ are two arbitrary complex numbers and $\Psi(x)$ is a
function whose explicit form will be given later. Having seen that
Eq.~(\ref{Weqn}) has a solution for $k=1$, let us address the question of
whether it also has a solution for all values of $k$. For that purpose one
has to investigate the adjoint of $\overline{D}$, and see whether its kernel
is null or not. Indeed, it is easy to see that the kernel of the adjoint vanishes, and hence 
Eq.~(\ref{Weqn}) has a regular solution no matter
what the left-hand side is, provided it is regular. Explicitly, one such 
solution is given by:
\bea
& W^{(k)} = D\,  U^{(k)} \\
\label{Udef}
\mbox{where  }\  & \overline{D} D\,   U^{(k)} = 
\imath\, \displaystyle{\sum_{l=1}^{k-1} \left(S^{(k-l)\dagger}W^{(l)} - W_c^{(l)\dagger}S^{(k-l)}\right)} \ .
\eea
The existence of a unique solution $U^{(k)}$ to Eq.~(\ref{Udef}) follows from 
the invertibility of the operator $\overline{D} D$. In terms of this
particular solution, and using
(\ref{W0eqn})-(\ref{W0sol}),   the most general solution of 
Eq.~(\ref{Weqn}) is given by:
\be
\label{arbitr}
W^{ (k)} = D\, U^{(k)} + \Psi(x)\, \left( \begin{array}{c c} K^{(k)}\ & Q^{(k)} \\ 0 \ & 0
\end{array} \right)\quad ,  
\ee
where  $K^{(k)}$ and $Q^{(k)}$ are complex constants.

Now let us study the solution of equation ~(\ref{Seqn}) order by order in
$\lambda$. It is easy to see
that both the left as the right hand sides are periodic functions in the
box. They can hence be expanded in Fourier series. The solution can be
obtained by equating the corresponding Fourier coefficients. However, notice
that the left-hand side has no constant term. Hence, if the right-hand has a
non-zero constant Fourier term, the equation has no solution. This can be expressed
more formally by saying that the kernel of $\partial$
is nontrivial.
What we will now show is that in solving the equation
for $S^{(k)}$ it is possible
to fix the constants  $K^{(k)}$ and $Q^{(k)}$ appearing in the solution to
(\ref{Weqn}) to order $k$,   by the condition that the
lowest Fourier component of the right-hand side vanishes.
For that we have
to explore the effect of  the replacement~(\ref{arbitr}) in Eq.~(\ref{Seqn})
to order $(\lambda\, \Delta)^{k}$. The dependence of the right-hand side of
Eq.~(\ref{Seqn}) on $K^{(k)}$ and $Q^{(k)}$ to this order, is contained in
the following term:
\be
 |\Psi(x)|^2\ ( c_3   \tau_3 + c_+   \tau_{+} + c_+^*   \tau_{-})\quad ,
\ee
where $c_3=2 \Re(Q^{(k)}Q^{(1)\, *}-K^{(k)}K^{(1)\,*})$ and
$c_+=2 (K^{(k)\, *}Q^{(1)}+K^{(1)\, *}\, Q^{(k)})$. The symbol $\Re$ denotes
the real part of its complex argument. By choosing $K^{(k)}$, $Q^{(k)}$
appropriately the term  within  parenthesis can be made equal to an arbitrary
hermitian, traceless $2 \times 2$ matrix.
Since the lowest Fourier coefficient
(the constant one) of $|\Psi(x)|^2$ is non-zero,  the constants
can be chosen such that the whole right  hand side of Eq.~(\ref{Seqn})
has vanishing constant Fourier term. Actually, this fixes 3 of the 4 real
parameters which enter  $K^{(k)}$, $Q^{(k)}$. The remaining one corresponds 
to  the symmetry associated to global colour rotations in the $1-2$ plane,
which leave $B_{\mu}$ invariant (we will comment upon this property on the next paragraph).  

Having shown that the solution of our set of equations exists order by order
in our expansion in $\sqrt{\lambda\Delta}$, we have now to analyse uniqueness.  Indeed,
on general grounds we know that the solution is non-unique. This fact is
associated to the existence of transformations which change one solution
into other. We already mentioned one: global gauge transformations of a
certain kind. These are the  residual gauge transformations that are not gauge fixed 
by Eq.~(\ref{gfcond}). They are associated with the freedom to multiply any solution
matrix $W$ by a constant phase. The other transformations are space-time 
translations (followed by an appropriate gauge transformation to preserve the gauge
fixing condition). This latter symmetry manifests itself under the form of a non-uniqueness
for the solutions of Eq.~(\ref{Seqn}) to any order $k$: notice that we are free
to add an arbitrary constant matrix to $S$ in the left-hand side of the
equation, which would entail the four real parameters associated to a translation.

The best strategy in solving the equations is to fix a unique solution 
by constraining these transformations. This we will do by imposing the 
following  additional
conditions:
\bea
\label{conda}
& \Re(W_{12}(x=0))=|W_{12}(x=0)|  \\
\label{condb}
& \int dx\,  S(x)=0 \quad ,
\eea
(If $W_{12}(x=0)=0$ we take $\Re(W_{11}(x=0))=|W_{11}(x=0)|$).
It is now completely clear that the procedure leads to a unique solution
order by order in $\sqrt{\lambda}$. 

It is useful to derive  expressions  for the field strength tensor itself.
Just as we did for the vector potential we might expand in colour components:
\be
 \Fmunu\, =  F_{\mu \nu}^{(3)}(x)\,   \tau_3 +F_{\mu \nu}^{(+)}(x)\,   \tau_+
 + F_{\mu \nu}^{*\, (+)}(x)\,   \tau_{-}\quad .
\ee
Now since the field is self-dual, we might contract it with the matrices
$ \sigma_{\mu} \overline{\sigma}_{\nu}$ to obtain a traceless 
hermitian matrix combining the three spatial directions:
\bea
\nonumber
{\cal F}^{(+)} &\equiv& \frac{-\imath}{4} F_{\mu \nu}^{ (+)}(x)\, 
\sigma_{\mu} \overline{\sigma}_{\nu} \\
\label{Fscriptplus}
&=& \frac{-\imath}{2}  D  W_c^{\dagger} -\half (S\, W_c^{\dagger} -W\,  S^{\dagger})  \\
\nonumber
{\cal F}^{(3)} &\equiv& \frac{-\imath}{4} F_{\mu \nu}^{(3)}(x)\,
\sigma_{\mu}  \overline{\sigma}_{\nu} \\
\label{Fscriptthree}
&=& \frac{\pi}{2}\left(\frac{l_0 l_3+ l_1 l_2}{V}\right)\tau_3\, -  \frac{\imath}{2}
\partial S^{\dagger} - \frac{1}{4} ( W\, W^{\dagger} -
W_c\, W_c^{\dagger})\quad .
\eea

Having set up the full procedure for calculating the potentials and fields in
powers of $\lambda$, let us now exemplify it by computing the first terms in 
this expansion. These results will be used in the next section. 
The starting point
is the equation for $W^{(1)}$ (Eq.~(\ref{W0eqn})). We mentioned previously
what is the form of the solution. Let us for the moment skip
the proof and also the determination of $\Psi(x)$ and proceed. 
The next step is to look at  the equation for $S^{(1)}$.
As mentioned in the general case, both the left and right-hand sides can be
expanded in Fourier coefficients. The condition that the constant coefficient
of the right hand side vanishes imposes a constraint on $K^{(1)}$ and
$Q^{(1)}$:
\bea
& 
Q^{(1)\, *} K^{(1)}=0\\
&|Q^{ (1)}|^2-|K^{ (1)}|^2=\frac{2 \pi}{\sqrt{V}}\quad , 
\eea
where we have fixed the normalisation of $\Psi(x)$, such that its constant
Fourier coefficient is equal to one. The previous equations lead to
\be
K^{(1)}=0 \ ; \quad \ Q^{ (1)}= \frac{\sqrt{2 \pi}}{V^{\frac{1}{4}}}\quad ,
\ee
where we have used (\ref{conda}).

Now the equation for $S^{(1)}$ reads:
\be
\overline{\partial} S^{(1)}= \frac{\imath \pi}{\sqrt{V }}(|\Psi(x)|^2-1)
\tau_3\quad .
\ee
This can be solved together with Eq.~(\ref{condb}) to give:
\be
\label{S1sol}
S^{(1)}= \frac{\imath \pi}{\sqrt{V}} (\partial h)   \tau_3\quad ,
\ee
where $h(x)$ is a periodic function on the box, solution of 
the  equation:
\be
\label{hdef}
\Box h(x)  = |\Psi(x)|^2-1  
\ee
and $\Box$ is the 4-dimensional Laplacian.
The previous equation can be solved by expanding both sides
in Fourier series  and equating.

Let us now work out the details of the solution to Eq.~(\ref{W0eqn}).
For future purposes we will consider a more general equation:
\be
\label{DQeqn}
\overline{D_q}\, \varphi(x) =0\quad ,
\ee
 where $\varphi$ is a two component vector. The operator $\overline{D_q}$
is given by:
\be
\label{DQdef}
\overline{D_q}=\overline{\sigma}_{\mu} (\partial_{\mu} + \imath \pi\, q
\frac{x_{\nu}\nmunu}{l_{\mu}l_{\nu}})\quad , 
\ee
where $q$ is a constant. The vector of functions   $\varphi(x)$ is required 
to satisfy the boundary condition:
\be
\label{qbc}
\varphi(x+ e_{\mu}) = \exp\{\imath \pi\,q\,  \nmunu \frac{x_{\nu}}{l_{\nu}}\}\;
\varphi(x)\, 
\ee
which is only consistent for integer $q$.
It is easy to see that the  operator  $\overline{D_q}$ preserves this boundary condition.

Now it is seen that locally a  solution of Eq.~(\ref{DQeqn}) takes the form:
\be
 \left( \begin{array}{c} \widetilde{\varphi}_q(x)\, \kappa_{+}(u_0,u_1) \\
(\widetilde{\varphi}_q(x))^{-1}\, \kappa_{-}(u_0^*,u_1^*)
\end{array}\right)\quad , 
\ee
where we have introduced complex coordinates:
\be
u_{\mu}=\frac{1}{l_{\mu}}(x_{\mu}+\imath\, n_{\mu \nu} x_{\nu})
\ee
and $u_{\mu}^*$ are the complex conjugates. These  coordinates are not 
independent, and satisfy:
\be
u_{\mu}= \frac {\imath}{l_{\mu}}  n_{\mu \nu} u_{\nu}l_{\nu} \quad .
\ee
We might for future benefit introduce the complex constants:
\be
\tau_{\mu}=  \frac {\imath}{l_{\mu}}  |n_{\mu \nu} l_{\nu}| \quad .
\ee

The function $\widetilde{\varphi}_q$ is given by:
\be
\widetilde{\varphi}_q(x)= \exp\{- \frac{\pi q}{2 l_0 l_3} (x_0^2+x_3^2) -
 \frac{\pi q}{2 l_1 l_2} (x_1^2+x_2^2)\}\quad .
\ee
The boundary conditions Eq.~(\ref{qbc}) impose constraints on the value
on the holomorphic and anti-holomorphic functions $ \kappa_{\pm}$:
\bea
&&  \kappa_{+}(x + e_{\mu})= \exp\{\pi q \imath  \frac{(u_{\mu} +\frac{1}{2})}{\tau_{\mu}}\}\  \kappa_{+}(x)\\
&&  \kappa_{-}(x + e_{\mu})= \exp\{-\pi q \imath \frac{(u_{\mu}^* + \frac{1}{2}    )}{\tau_{\mu}
}\}\ \kappa_{-}(x)\quad . \eea
Choosing $u_0$ and $u_1$ as our two independent complex variables, we might write:
\bea
&&\kappa_{+}(u_0,u_1) = \exp \{ \pi q \imath (\frac{u_0^2 }{2\tau_0} +
\frac{u_1^2 }{2\tau_1})  \}\  \widetilde{\kappa}_{+}(u_0,u_1)\\
&&\kappa_{-}(u_0^*,u_1^*) = \exp \{ -\pi q \imath (\frac{u_0^{* 2 }}{2\tau_0} +
\frac{u_1^{ * 2 }}{2\tau_1})  \}\  \widetilde{\kappa}_{-}(u_0^*,u_1^*)
\eea 
The functions $\widetilde{\kappa}_{\pm}$ are periodic in their arguments with period $1$ and satisfy:
\bea
&&\widetilde{\kappa}_{+}(u_0+\tau_0,u_1)= \exp\{-2 \pi q \imath u_0 -\pi q \imath \tau_0\}\  \widetilde{\kappa}_{+}(u_0,u_1) \\
&&\widetilde{\kappa}_{+}(u_0,u_1+\tau_1)= \exp\{-2 \pi q \imath u_1 -\pi q
\imath \tau_1\}\  \widetilde{\kappa}_{+}(u_0,u_1)\quad .
\eea
For $q=1$ these are precisely the conditions satisfied by the  Riemann $\theta$
function~\cite{theta}. Actually, up to a multiplicative constant, this function is 
the only (regular) holomorphic function satisfying these boundary conditions.
 Similarly, one obtains that the equation for $\widetilde{\kappa}_{-}$ has no
 regular  solutions.
Hence, for the $q=1$ case we have arrived at the solution given  in
 Eq.~(\ref{W0sol}), and determined the expression for  the function $\Psi(x)$:
\be
\label{psiexp}
\Psi(x)= \sqrt{\frac{4 l_3 l_2}{l_0 l_1}}\;  \exp\{ -\frac{\pi }{l_0 l_3} (x_3^2 -\imath x_3 x_0) -\frac{\pi }{l_1 l_2} (x_2^2 -\imath x_2 x_1)  \}\; \theta(u_0,\tau_0)\, \theta(u_1,\tau_1)
\ee
The multiplicative factor  preceding the right hand side  of the previous expression 
is determined by the condition that  the lowest Fourier coefficient 
of $|\Psi(x)|^2$ is unity. 

%\input{comparison}

%\input{Nahm}

%\end{document}

\section{Comparison with numerical results}

In the previous section we have set up a general procedure to construct the form of the gauge potentials
and field strengths for $\SUT$ self-dual solutions on the torus with twist tensor $n_{03}=n_{12}=1$.
The result is an expansion in powers of $\sqrt{\Delta}$, where  $\Delta$ is defined in Eq.~(\ref{Deltadef}).
However, we do not
have an analytical estimate of the size of the coefficient. Our purpose in
this section is to test the rate of convergence of the expansion by
comparing the results obtained from the first non-trivial order with 
the  exact result as obtained by numerical methods on the lattice.

We will restrict to the analysis of the gauge invariant traces $\mbox{Tr}(\mathbf{E}_i\mathbf{E}_j)$,
where $\mathbf{E}_i=\mathbf{F}_{0i}$ are the electric fields. To lowest order in our
expansion, we fall into the constant field strength configuration, and the
only non-zero gauge invariant trace is $\mbox{Tr}(\mathbf{E}_3^2)$.
The next correction is order $\Delta$ and vanishes for
$\mbox{Tr}(\mathbf{E}_1 \mathbf{E}_2)$.  It also predicts that 
 $\mbox{Tr}(\mathbf{E}_1^2) =
\mbox{Tr}(\mathbf{E}_2^2)$. 
Making use of the 
general formulas~(\ref{Fscriptplus}),~(\ref{Fscriptthree}) and substituting 
the explicit form of $S^{(1)}$ and
$W^{(0)}$ (Eqs.~(\ref{S1sol}),~(\ref{W0sol})) one arrives
at the following result, valid  to order $\Delta$:
{\setlength\arraycolsep{2pt}
\bea
\label{Trnonab}
\mbox{Tr}(\mathbf{E}_1^2(x)) &=& \mbox{Tr}(\mathbf{E}_2^2(x)) =
\Delta \frac{\pi}{\sqrt{V}}\left|D_0\Psi(x)\right|^2 \ , \\
\nonumber
\mbox{Tr}(\mathbf{E}_3^2(x)) &=& \frac{\pi^2}{2}\left(\frac{l_0l_3+l_1l_2}{V}\right)^2 \times \\
\label{Trab}
&& \relax{\kern-0mm}\left\{
1-\Delta\left(\frac{2\sqrt{V}}{l_0l_3+l_1l_2}\right)\left(1+2(\partial_0^2+\partial_3^2)h(x)\right)
\right\} \ , \\
\label{Tmix13}
\mbox{Tr}(\mathbf{E}_1(x) \mathbf{E}_3(x)) &=& -\Delta \pi^2 \left(\frac{l_0l_3+l_1l_2}{V^{3/2}}\right)
\left(\partial_0\partial_2+\partial_1\partial_3\right)h(x) \ , \\
\label{Tmix23}
\mbox{Tr}(\mathbf{E}_2(x) \mathbf{E}_3(x)) &=& \Delta \pi^2 \left(\frac{l_0l_3+l_1l_2}{V^{3/2}}\right)
\left(\partial_0\partial_1-\partial_2\partial_3\right)h(x) \ ,
\eea
}
where $\Psi$ and $h$ have been defined in Section 2
(Eqs.~(\ref{psiexp}) and~(\ref{hdef})). Using the standard representation of Riemann's 
theta function~\cite{theta}:
\be
\theta(u,\tau)=\sum_{n\in {\mathbf Z}} \exp\{ 2 \pi \imath n u + \pi \imath
n^2 \tau\}  
\ee
one can easily obtain the Fourier coefficients of all the functions appearing in
Eqs.~(\ref{Trnonab}-\ref{Tmix23}). As for the numerical comparison, summing the first few hundred terms of the
Fourier expansion allows to compute these functions with negligible errors. It is also extremely simple
to use these Fourier coefficients to integrate analytically over some of the four real
coordinates, to arrive at a quantity better suited for graphically displaying the
comparison.

A numerical approximation to the exact solutions of the self-duality equations, with which the 
results coming from the perturbative 
approximation that
we have developed are to be compared, can be constructed  by means of
standard lattice gauge
theory techniques~\cite{Numerical,overimproved}. We will use for this purpose an $\varepsilon=0$ overimproved cooling 
procedure,
that was found in previous works~\cite{overimproved} to be able to produce very accurate 
approximants to continuum
self-dual fields. In particular, it allows to extract the exact values of the gauge invariant densities
$\mbox{Tr}(\mathbf{E}_i\mathbf{E}_j)$ under concern up to ${\cal O}(a^4)$ corrections, $a$ being the lattice spacing
(whose precise definition we will discuss below).

To explore the accuracy of the next to leading term  in the perturbative
expansion with varying values of the remaining parameters,  we will consider tori of lengths
$(l_0=l_t(1+\epsilon),l_1=l_t,l_2=l_s,l_3=l_s)$. The results will then
depend on  the perturbative parameter $\Delta=\epsilon/\sqrt{1+\epsilon}$,
and on the ratio $l_s/l_t$  measuring the degree of spatial asymmetry of the
torus. Thus, by keeping $0\le\epsilon\ll 1$ we will remain within the perturbative regime.
On the other hand, we will vary the ratio $l_s/l_t$ between $1$ (the more
symmetrical case) and $0$, were the $\theta$ functions and their derivatives
entering into the analytical expressions are well described by polynomials
times gaussians.

In each case, a numerical solution is obtained in a lattice with a number of points
$L_{\mu}=l_{\mu}/a$ along direction $\mu$.
To define the lattice spacing $a$, we need to fix a unit. In our case we
take $l_1 l_2 =1$. This is justified by  noticing  from the expression for $\Psi$
in Eq.~(\ref{psiexp}) that the region having
nontrivial structure in the action density  at nonzero $\Delta$
is of size $\sqrt{l_1 l_2}$.

We will present the results of the comparison of the analytical results  Eqs.~(\ref{Trnonab}-
\ref{Tmix23}) with their  numerical counterparts,
for three different configurations, having different values of $\Delta$ and $l_s/l_t$. 
The lattice sizes that we will use,
together with their associated $\Delta$ and $l_s/l_t$ values, are detailed in Table~\ref{tab:lattparams}.

\begin{table}[h!]
\begin{center}
\begin{tabular}{|c|c|c|c|}
\hline
Lattice & Size & $\Delta$ & $l_s/l_t$ \\
\hline\hline
A & $13 \times 12 \times 12 \times 12$ & 0.080064 & 1.00 \\ \hline
B & $21 \times 20 \times  8 \times  8$ & 0.048795 & 0.40 \\ \hline
C & $41 \times 40 \times  6 \times  6$ & 0.024693 & 0.15 \\ \hline
\end{tabular}
\end{center}
\caption{Lattices used in the comparison, and their associated $\Delta$ and $l_s/l_t$ values.}
\label{tab:lattparams}
\end{table}

\begin{figure}[htb]
\vspace{17.0cm}
\includegraphics{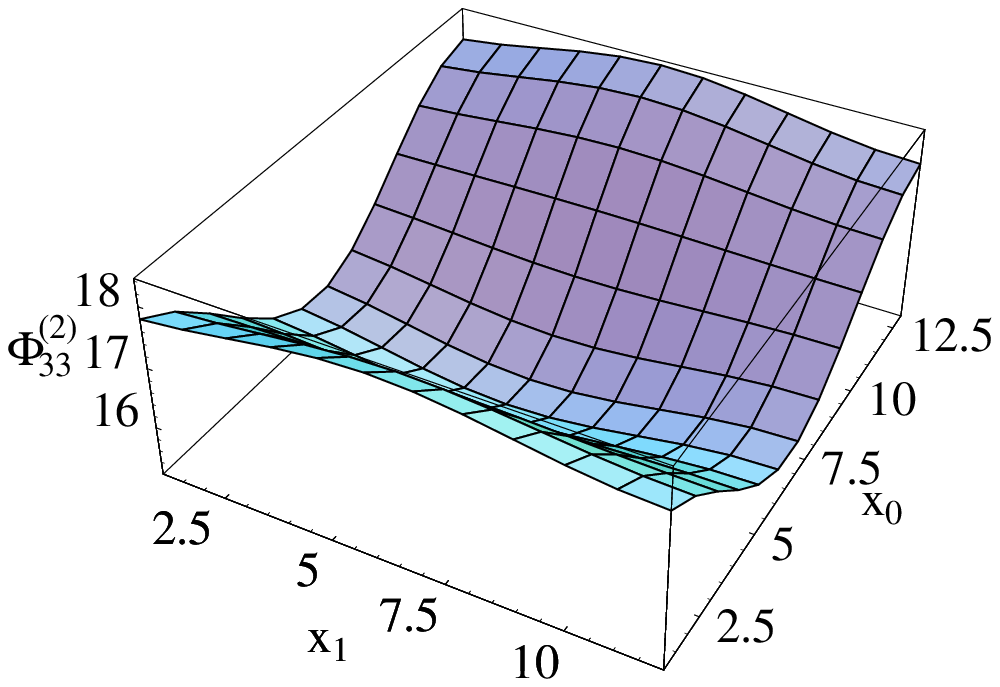}
\includegraphics{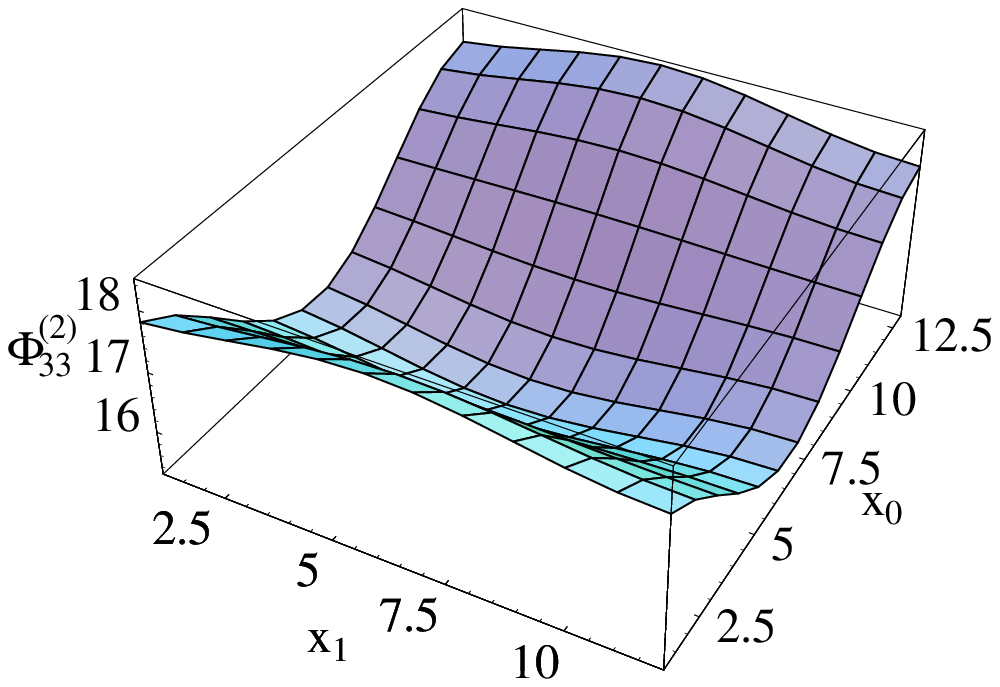}
\includegraphics{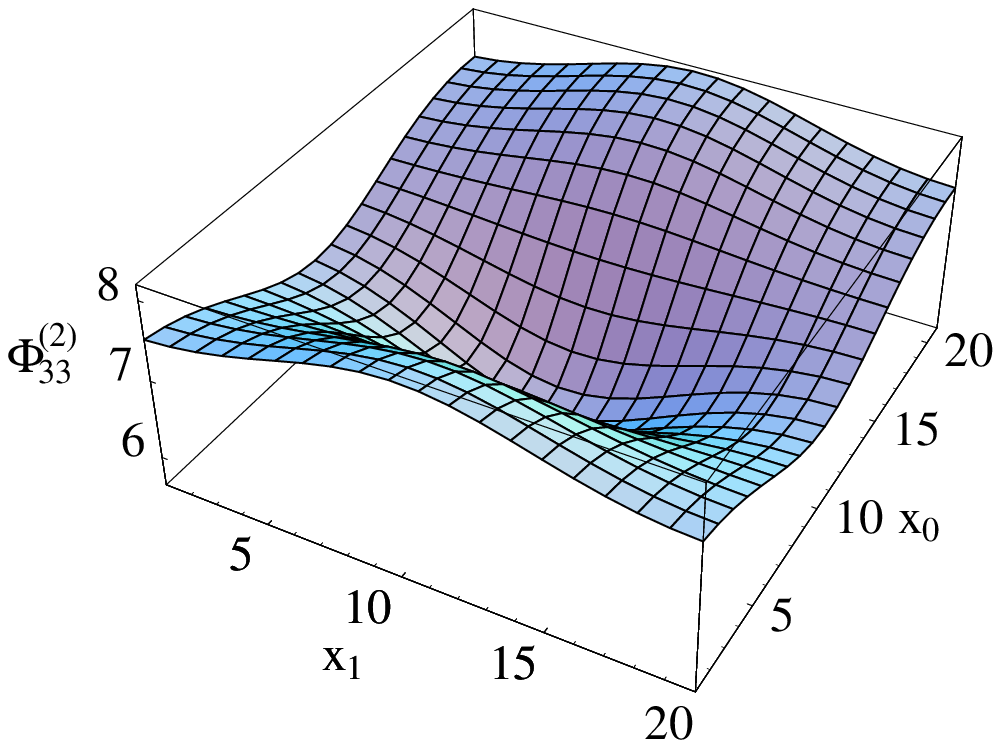}
\includegraphics{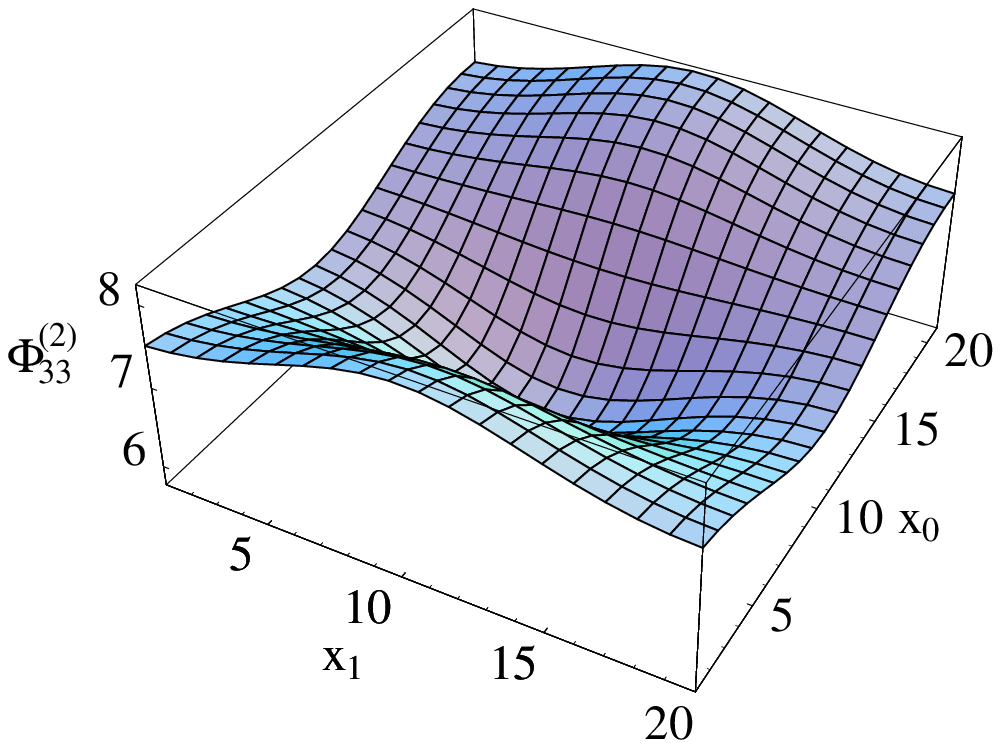}
\includegraphics{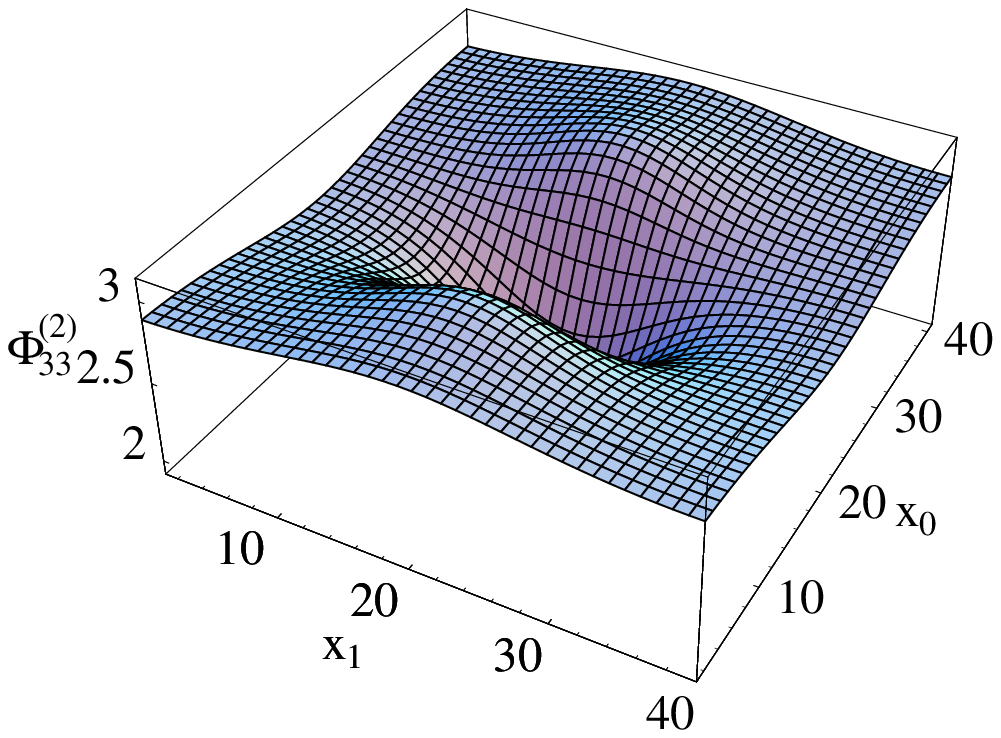}
\includegraphics{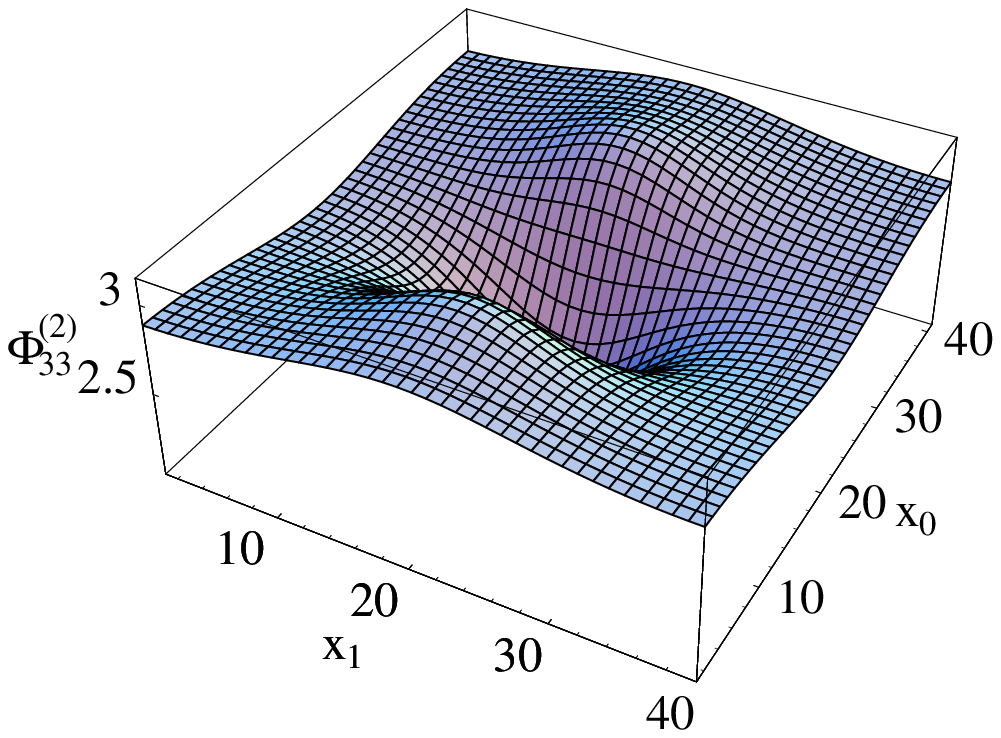}
\caption{Invariant densities $\Phi_{33}^{(2)}$ for the configurations A, B, C of Table~\ref{tab:lattparams},
from top to bottom, are shown. Plots in the left column display the analytical perturbative result,
and plots in the right column display the exact (numerical) result.}
\label{fig:twod_compt}
\end{figure}

\begin{figure}[htb]
\vspace{17.0cm}
\includegraphics{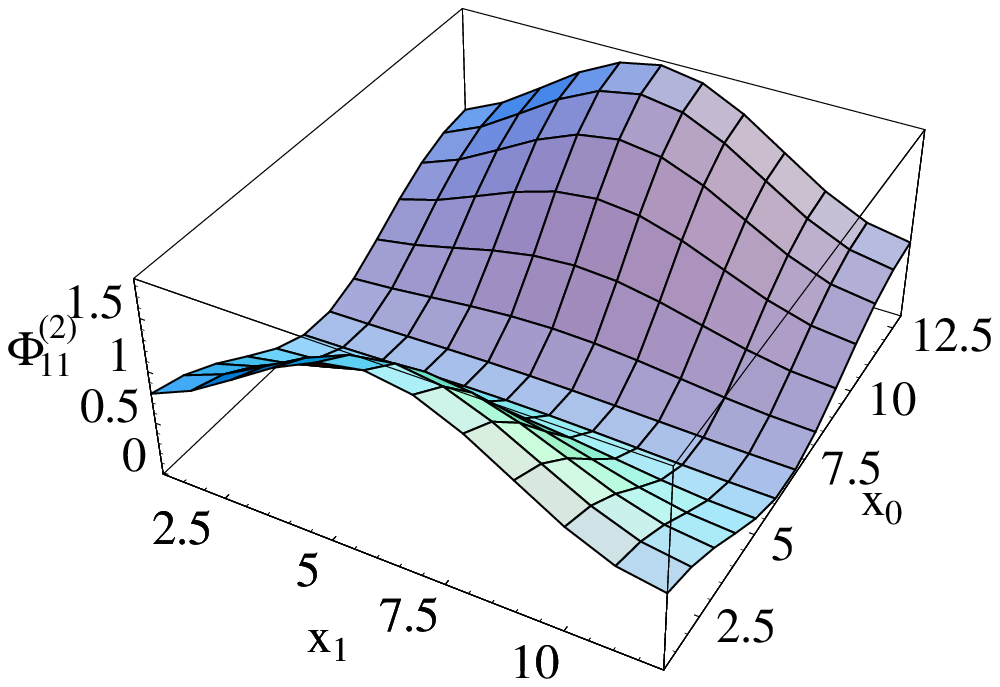}
\includegraphics{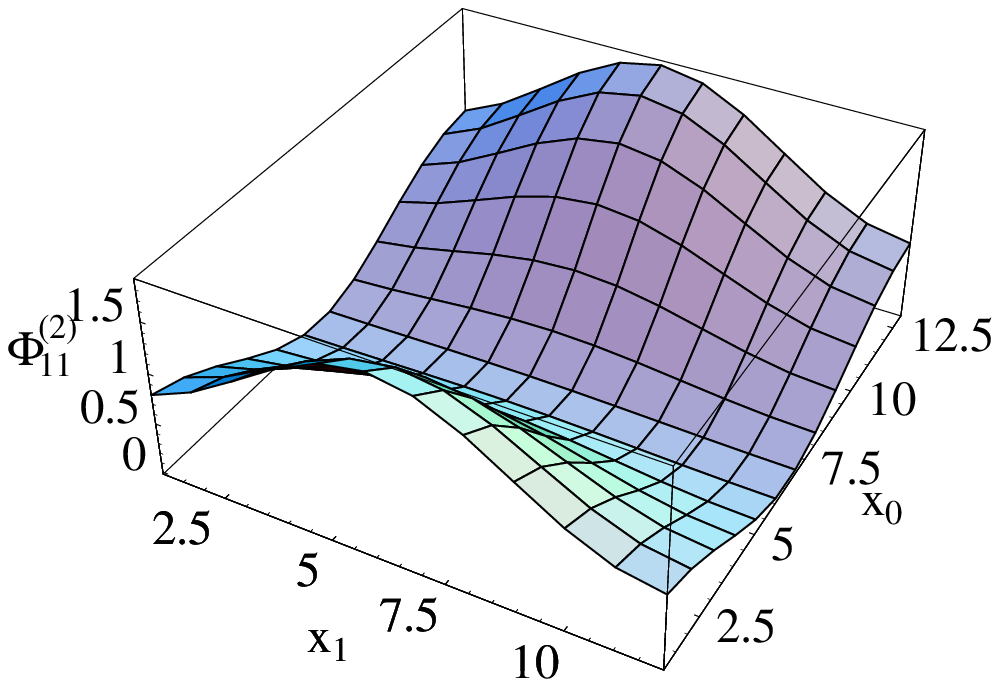}
\includegraphics{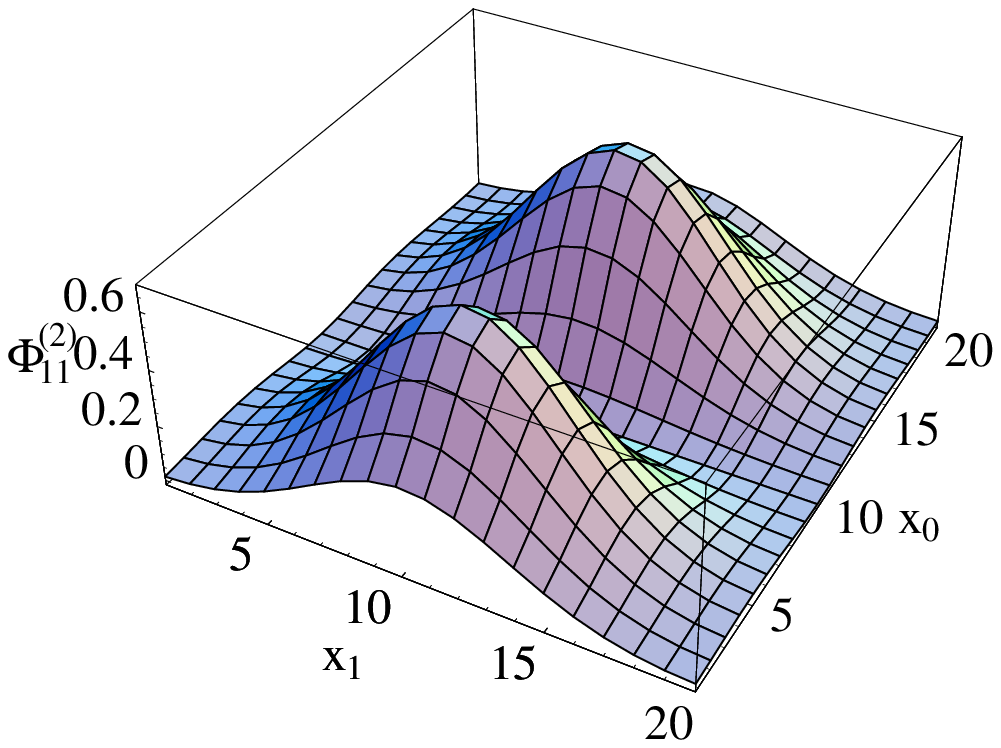}
\includegraphics{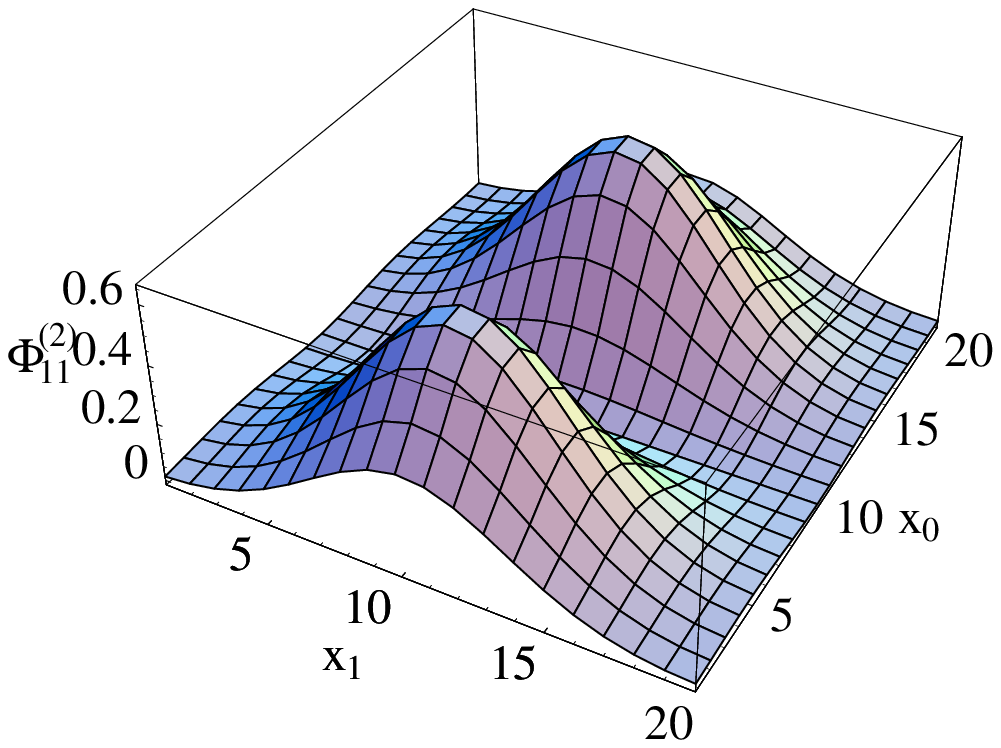}
\includegraphics{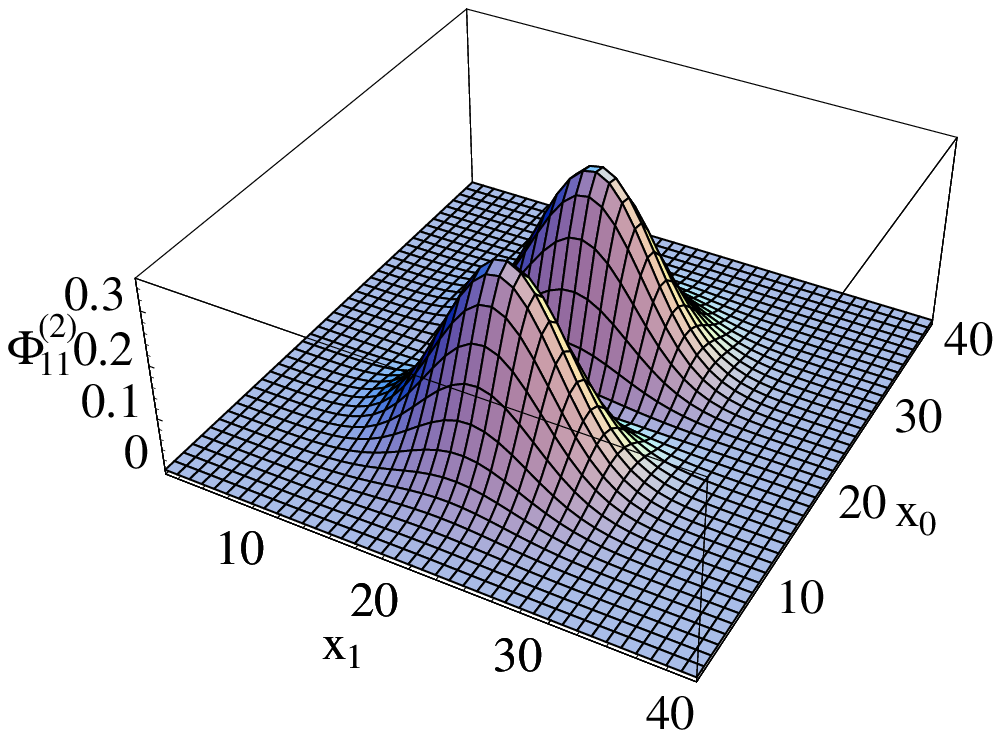}
\includegraphics{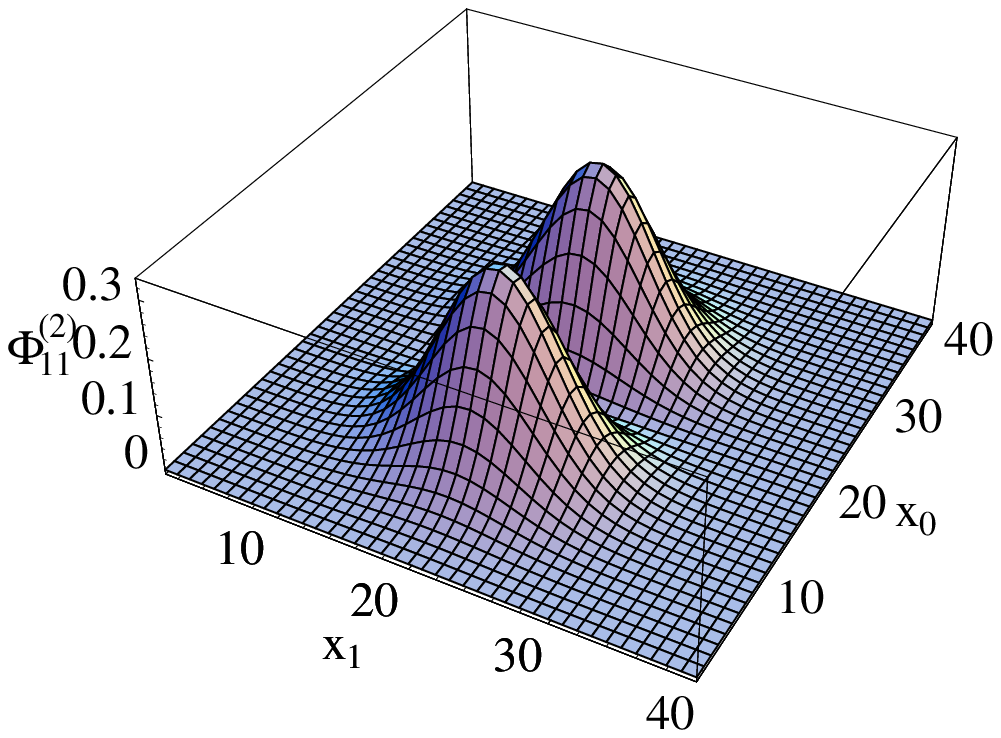}
\caption{Invariant densities $\Phi_{11}^{(2)}$ for the configurations A, B, C of Table~\ref{tab:lattparams},
from top to bottom, are shown. Plots in the left column display the analytical perturbative result,
and plots in the right column display the exact (numerical) result.}
\label{fig:twod_compp}
\end{figure}

\begin{figure}[htb]
\vspace{17.0cm}
\includegraphics{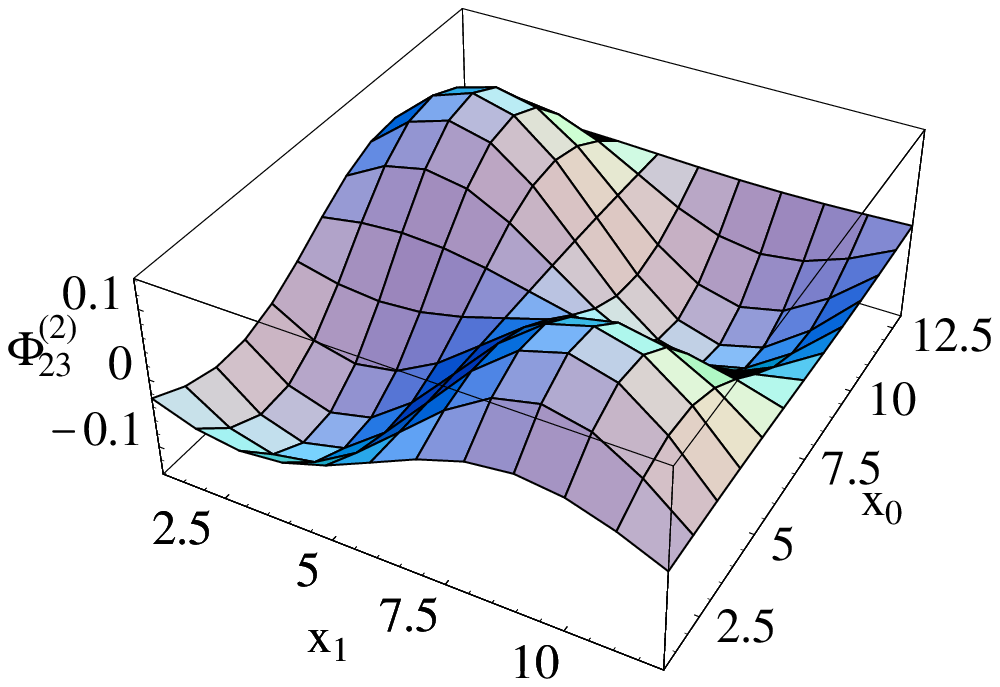}
\includegraphics{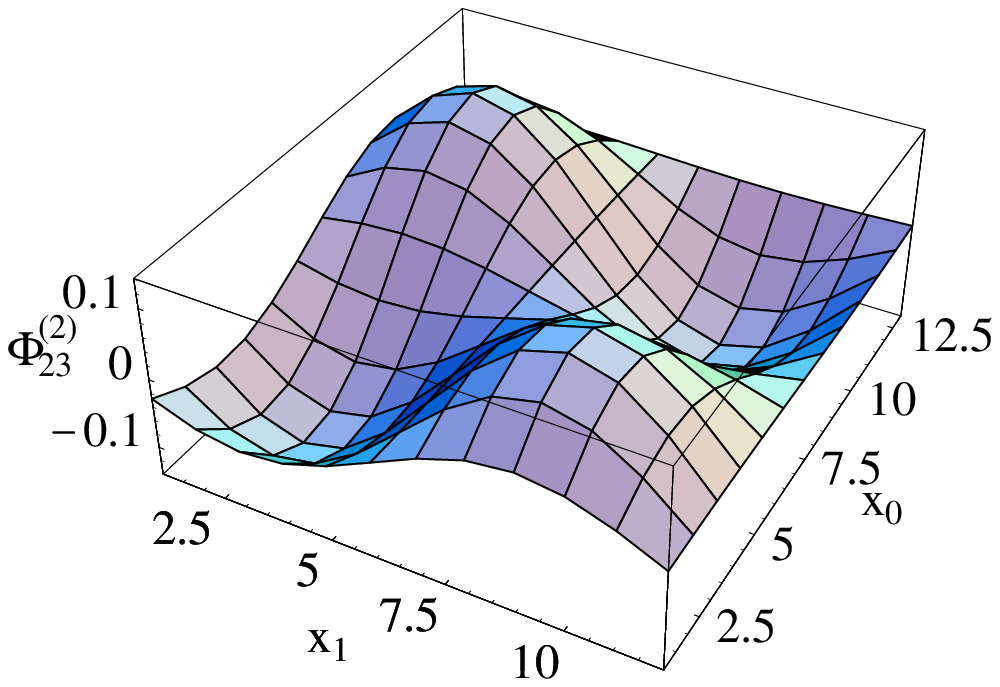}
\includegraphics{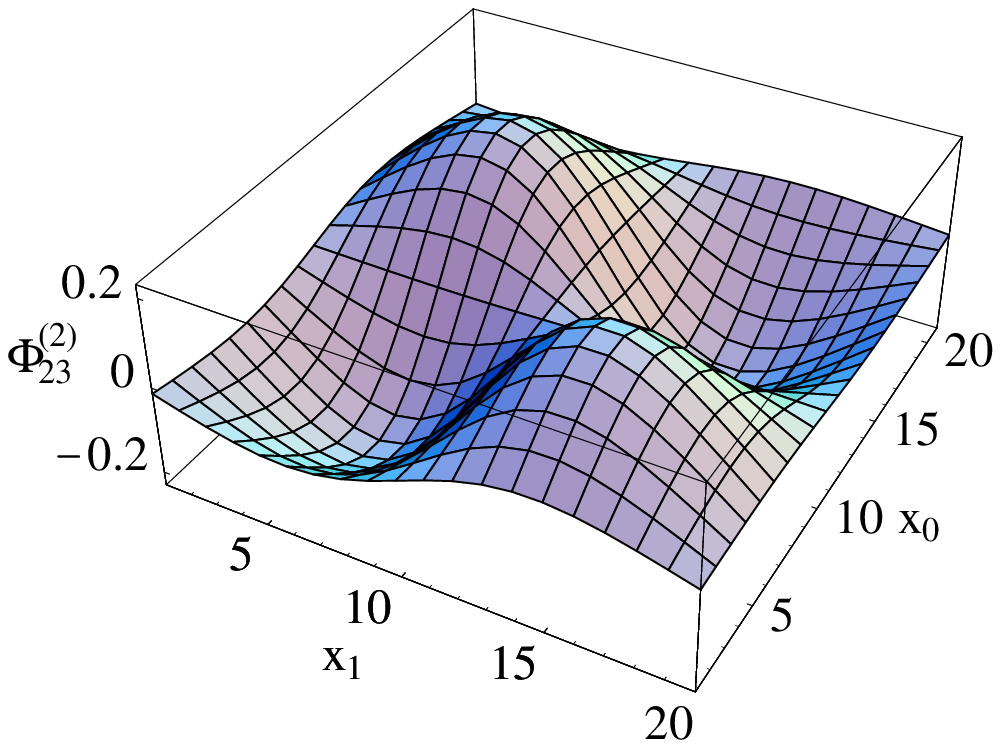}
\includegraphics{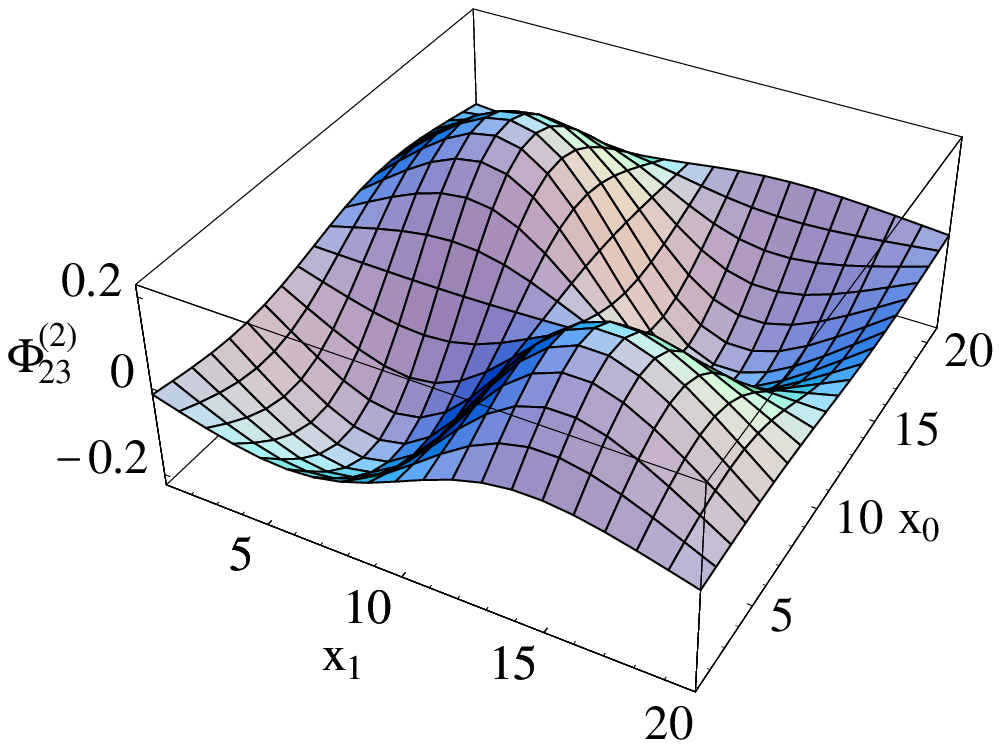}
\includegraphics{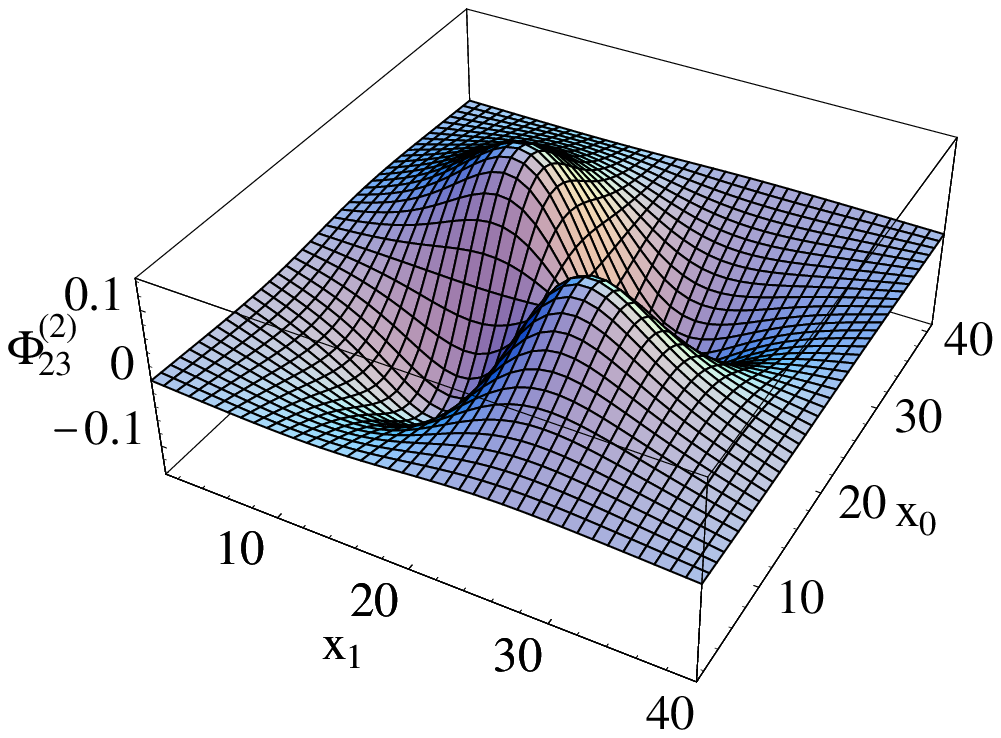}
\includegraphics{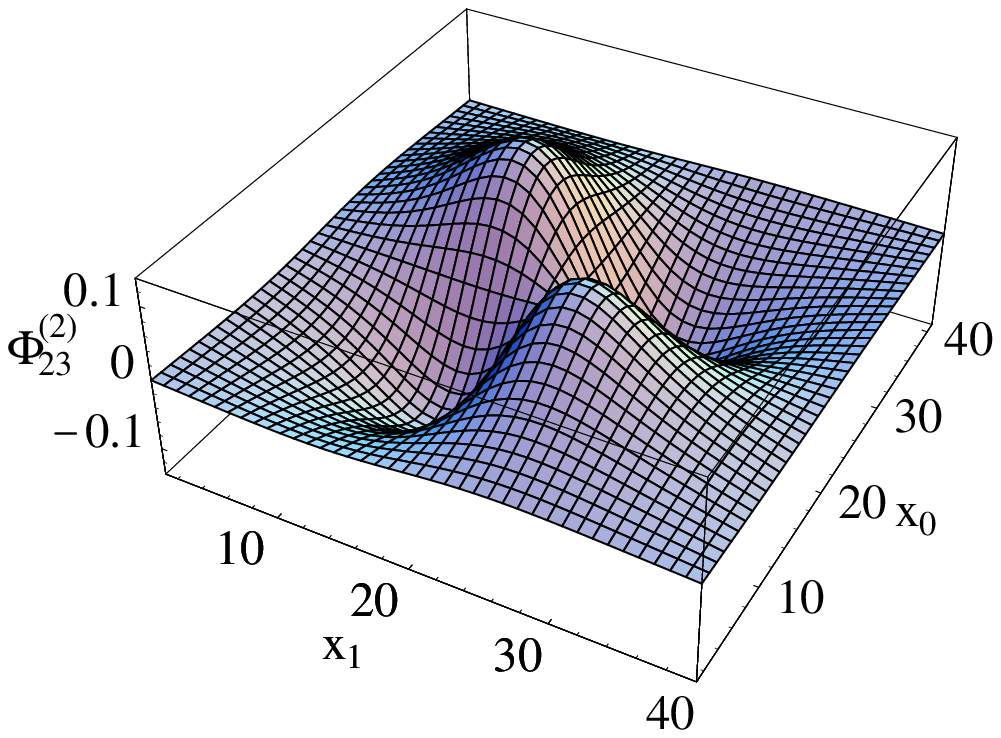}
\caption{Invariant densities $\Phi_{23}^{(2)}$ for the configurations A, B, C of Table~\ref{tab:lattparams},
from top to bottom, are shown. Plots in the left column display the analytical perturbative result,
and plots in the right column display the exact (numerical) result.}
\label{fig:twod_compmix}
\end{figure}

\begin{figure}[htb]
\vspace{15.5cm}
\includegraphics{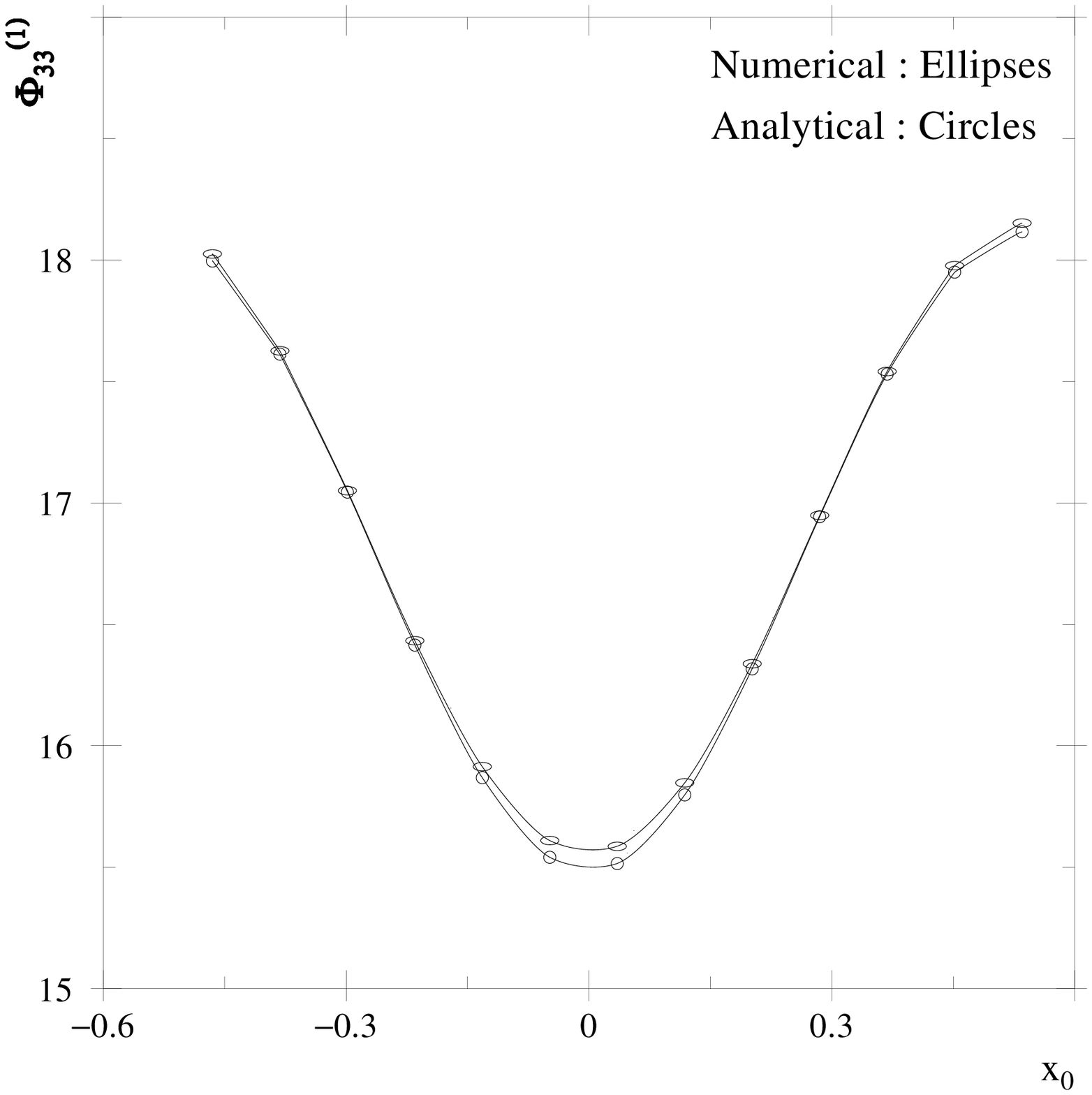}
\includegraphics{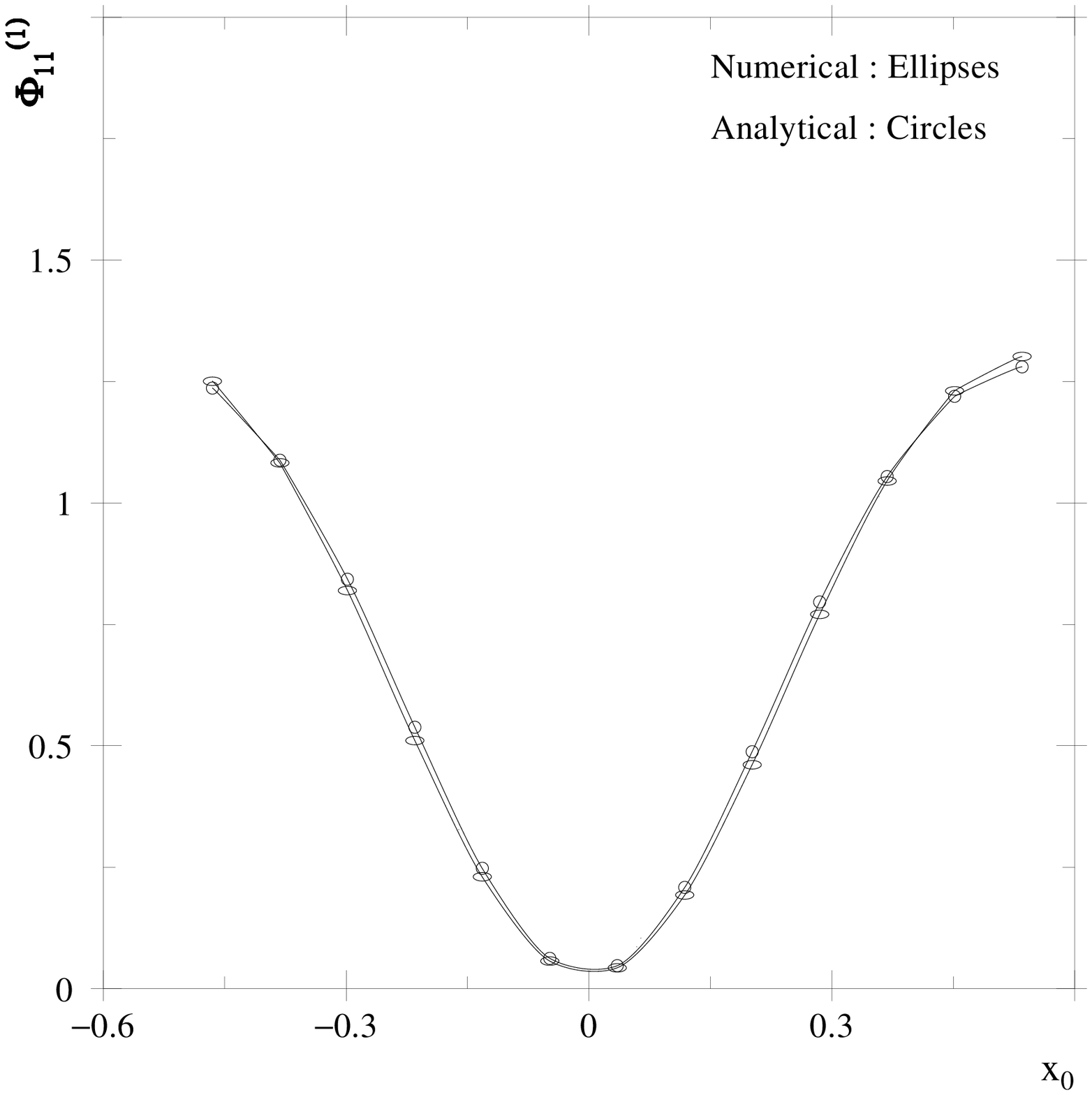}
\includegraphics{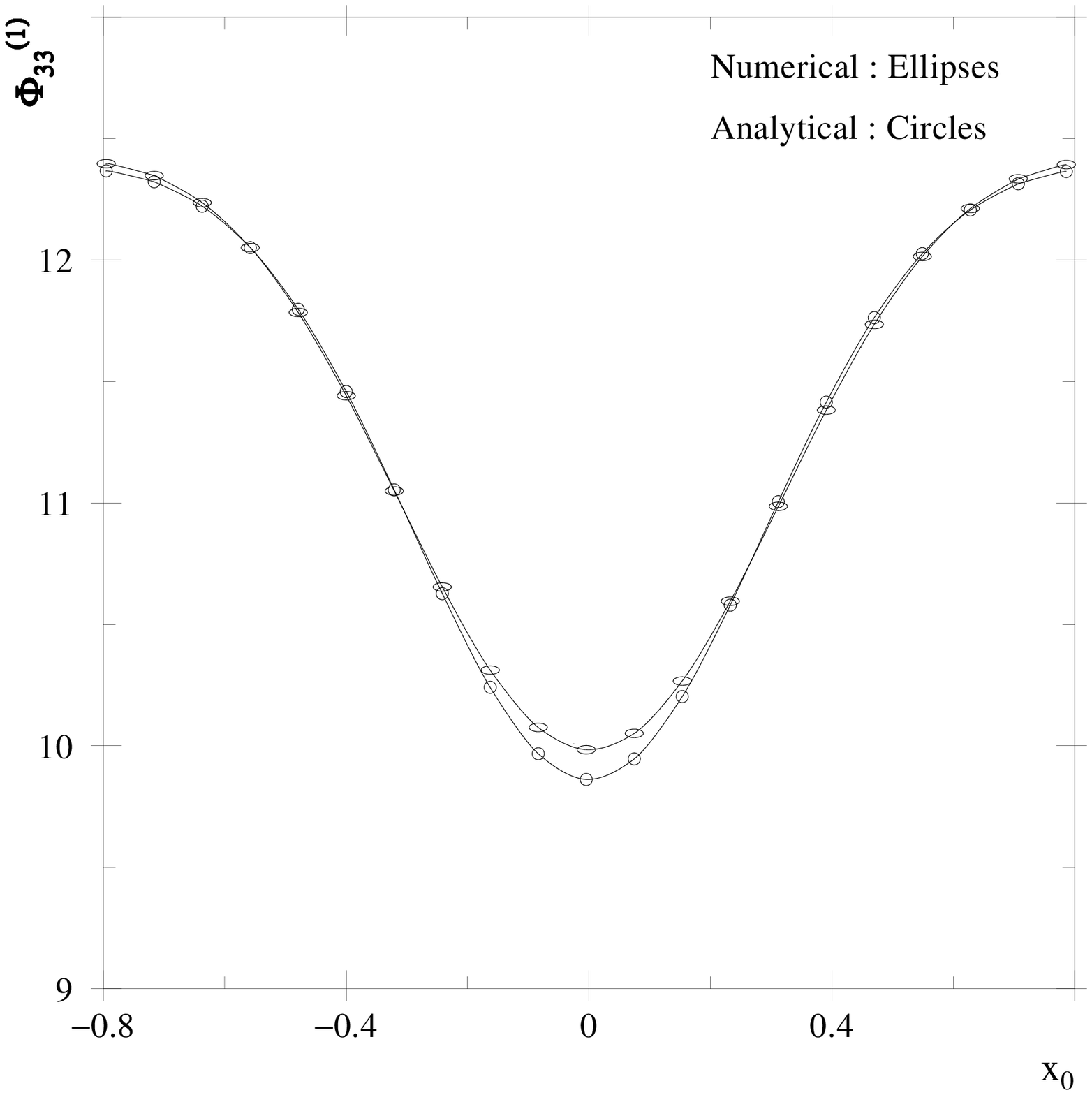}
\includegraphics{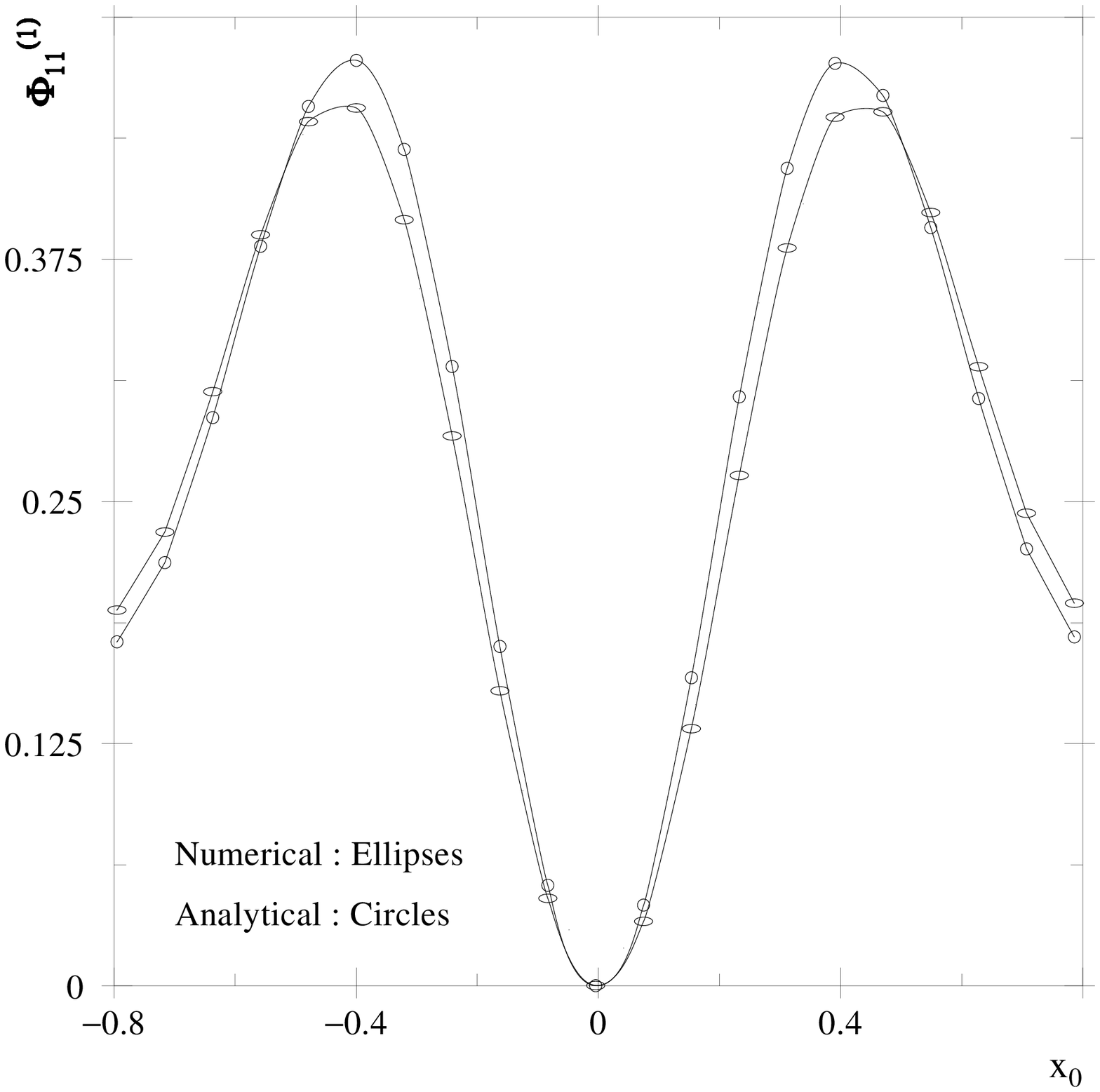}
\includegraphics{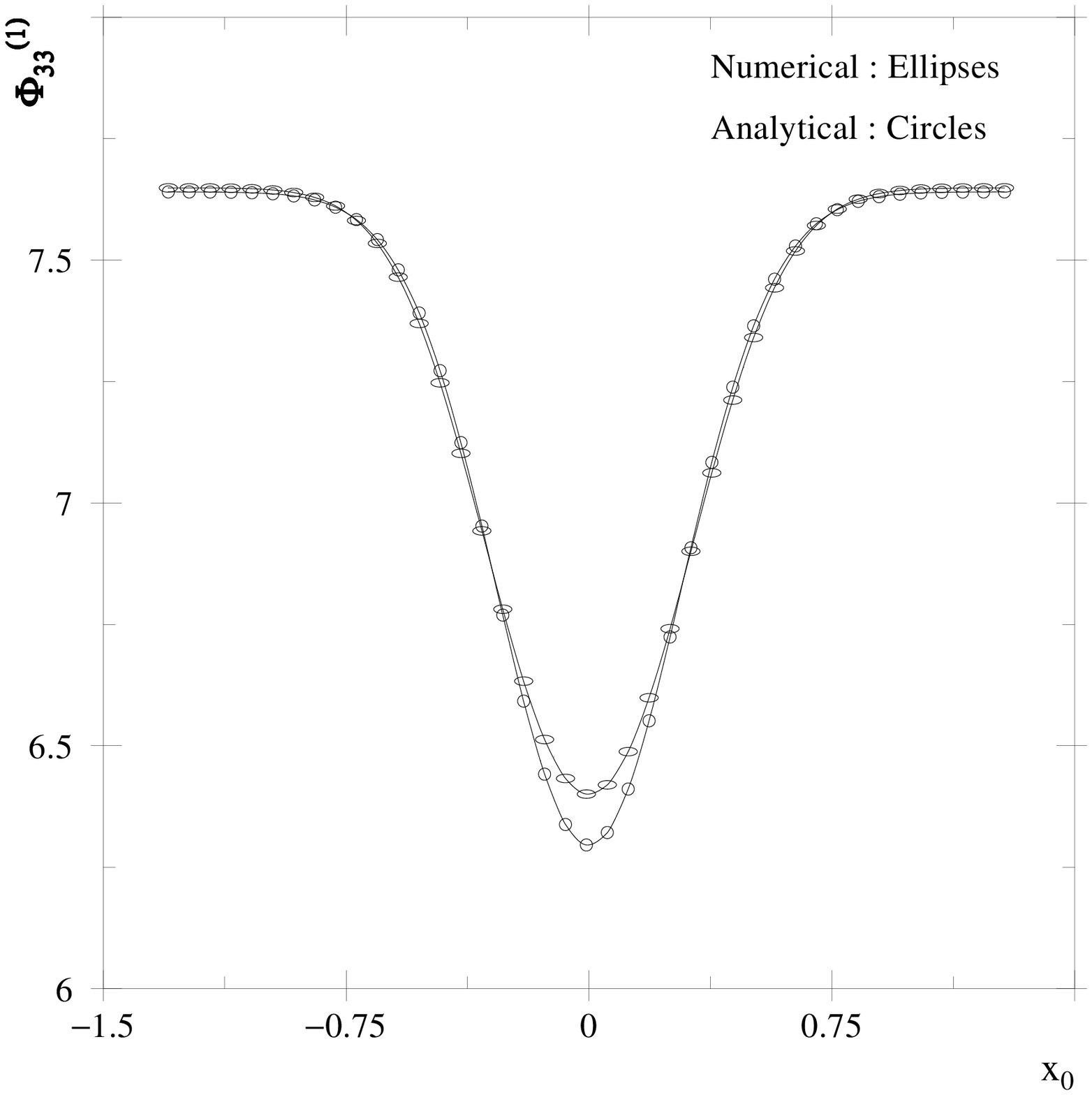}
\includegraphics{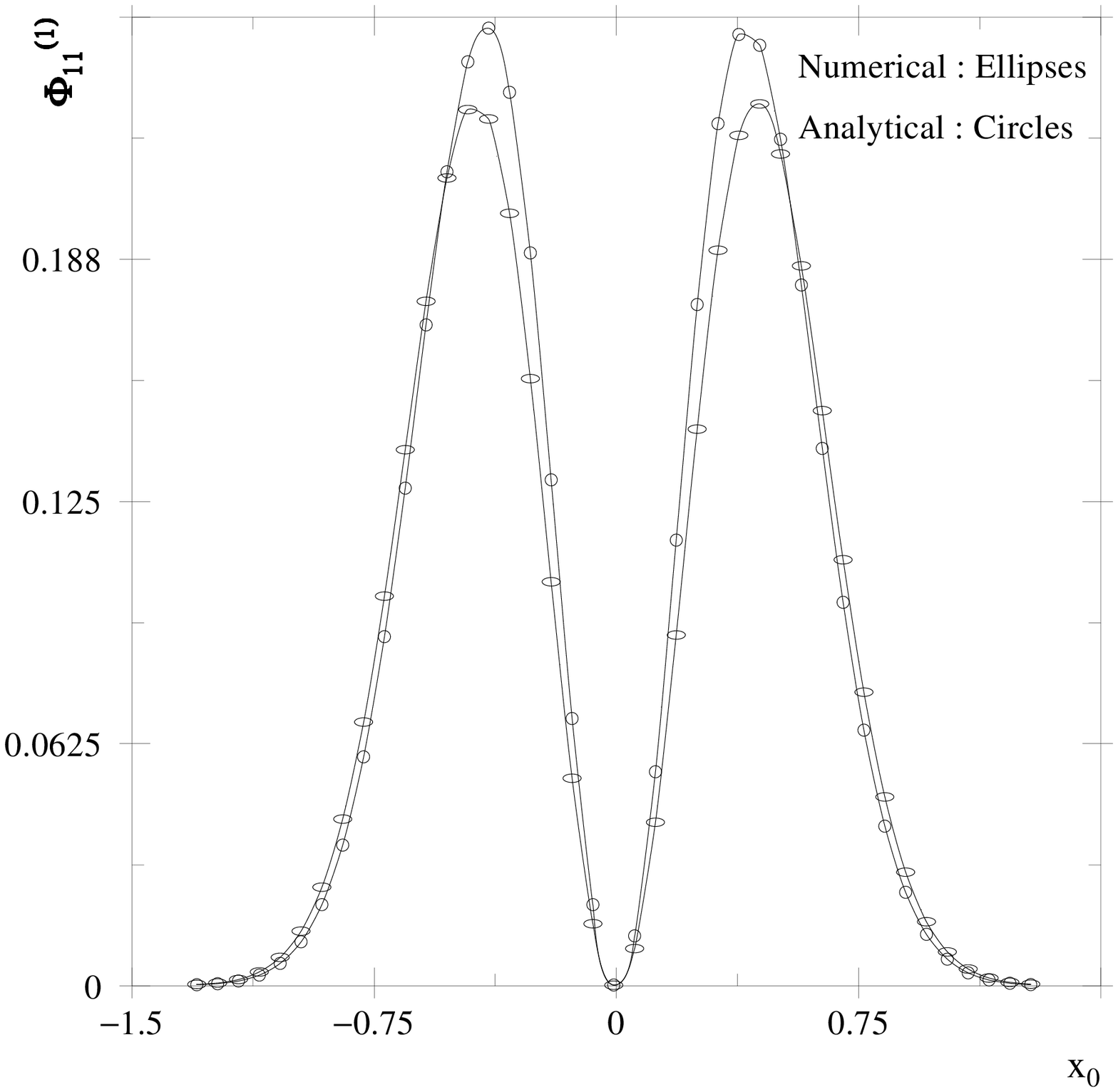}
\caption{Comparison of exact and perturbative results for the electric field Euclidian
time profiles $\Phi_{33}^{(1)}$ (left column) and $\Phi_{11}^{(1)}$ (right column) in the configurations
A, B, C, from top to bottom, of Table~\ref{tab:lattparams}. The $x_0$ coordinate has been rescaled
with the lattice spacing $a$ as defined in the text, and the center of the object has been set as
origin of coordinates. The interpolating lines are intended to guide the eye.}
\label{fig:oned}
\end{figure}

In Figures~\ref{fig:twod_compt},~\ref{fig:twod_compp} and~\ref{fig:twod_compmix} we
show the numerical and perturbative results for the
integrated electric field densities $\Phi_{33}^{(2)}(x_0,x_1)\equiv\int\mbox{d}x_2\mbox{d}x_3\mbox{Tr}(\mathbf{E}_3^2(x))$
and $\Phi_{11}^{(2)}(x_0,x_1)=\Phi_{22}^{(2)}(x_0,x_1)\equiv\int\mbox{d}x_2\mbox{d}x_3\mbox{Tr}(\mathbf{E}_1^2(x))$,
and $\Phi_{23}^{(2)}(x_0,x_1)\equiv\int\mbox{d}x_2\mbox{d}x_3\mbox{Tr}(\mathbf{E}_2(x)\mathbf{E}_3(x))$,
respectively (notice that $\Phi_{13}^{(2)}$ vanishes at the present perturbative order, despite the fact that
$\mbox{Tr}(\mathbf{E}_1(x)\mathbf{E}_3(x))$ does not, because of the particular form of the expression for
this latter quantity). The qualitative agreement is clearly good. The main features of the exact
solution are present in the analytical expression. It is possible to obtain a
graphical  quantitative measure  of the comparison by integrating the
previous densities over an additional coordinate to yield the  time profiles
$\Phi_{33}^{(1)}(x_0)\equiv\int\mbox{d}x_1\mbox{d}x_2\mbox{d}x_3\mbox{Tr}(\mathbf{E}_3^2(x))$
and
$\Phi_{11}^{(1)}(x_0)=\Phi_{22}^{(1)}(x_0)\equiv\int\mbox{d}x_1\mbox{d}x_2\mbox{d}x_3\mbox{Tr}(\mathbf{E}_1^2(x))$
(similarly to what happened with $\Phi_{13}^{(2)}$, $\Phi_{23}^{(1)}$ vanishes despite
$\Phi_{23}^{(2)}$ does not). The comparison for these quantities is displayed in Figure~\ref{fig:oned}.

Let us briefly comment some salient features of the solution. As for the density of the component
$\mathbf{E}_3$ of the electric field, a hole
appears overlaying the flat background supplied by the zero-order constant abelian field. The width of this
structure is, as we mentioned above, proportional to $\sqrt{l_s l_t}$, as can be derived from an analysis of the
perturbative expression for the potential, and its contribution to the total action is of order $\Delta$
(cf. Eq.~(\ref{Trab})). Meanwhile, the action density associated to the other components of the electric field, whose
contribution to the total action is as well of order $\Delta$ in the perturbative approach, exhibits in the asymmetric
torus case (configurations B and C) a double lump structure, again of size $\sqrt{l_s l_t}$, with the maxima aligned in Euclidian time. We see that the
perturbative approximation is most accurate for the configuration A, despite the relatively large value of $\Delta$
associated to it. This fact indicates that the convergence behaviour of the perturbative series depends on the value
of the asymmetry parameter $l_s/l_t$, in such a way that it is worse the more asymmetric the torus is chosen.
The overall conclusion is, anyway, that for values of $\Delta$ in the probed range, between $0.02$ and $0.09$,
the NLO perturbative result constitutes a good approximation to the exact solution.

Once the convergence behaviour of the perturbative series for small values of $\Delta$ has been checked to
be good, it would be also interesting to study to what extent the NLO result at hand remains useful to
describe solutions occurring at larger values of $\Delta$. This possibility is tempting because it would
open the door to apply our results to improve the analytical control over some particularly interesting
fields. For instance, it is known~\cite{Numerical,overimproved} that in a torus of geometry $l_t \times l_s^3$ with $l_s/l_t \ll 1$
the solutions approach self-dual fields on $T^3\times\Real$, the approximation being already remarkably good for
$l_t \sim 3l_s$. For the considered twist and $\SUT$ gauge group a $Q=1/2$ solution is obtained, whose action
density displays a single lump exponentially decaying in the large direction of length $l_t$, and whose width
is controlled by $l_s$. In this geometry, and setting $l_t \equiv l_s(1+\delta)$, one has $\Delta=\delta/\sqrt{1+\delta}$,
and the analysis would proceed by moving from the case $\delta=0$, where the torus is symmetric and the solution is
the abelian one, to values of $\delta \sim 1$, where the features of the $T^3\times\Real$ solution would
start to arise. Having performed computations on lattices of sizes $L_t \times L_s^3$, with $L_s=12$ and $L_t$
ranging from $13$ to $48$, we have found that the self-dual configuration evolves smoothly with changing
$\delta$. Unfortunately, the NLO perturbative approximation begins to deviate substantially from the exact result
before the interesting regime $\delta \gtrsim 1$ is reached.

In the same spirit one could investigate other torus geometries, e.g. $l_t^2 \times l_s^2$ with $l_s/l_t \ll 1$,
which in some cases is known to lead to limiting $T^2\times\Real^2$ solutions with a vortex-like structure~\cite{vortex},
or $l_t \times l_s^3$ with $l_s/l_t \gg 1$, which leads to the $\Real^3\times S^1$ caloron solutions~\cite{Calo}.
In these cases, we would expect a similar behaviour to that found for the $T^3\times\Real$ case.

\section{Nahm transform}

Nahm's transformation~\cite{Nahm,PBPvB} maps self-dual configurations on the torus 
onto other self-dual configurations. The modifications necessary to cope 
with twisted boundary conditions have only been worked out
recently~\cite{twist,Calod}. 
In general, the transformation changes  the twist tensor  and 
torus sizes and maps the rank of the group ($N$) and the topological charge
($Q$) 
onto each other through the formula:
\bea
 & Q \longrightarrow Q'= N/N_0 \\
 & N  \longrightarrow N'= Q N_0 \quad ,
\eea  
which preserves the dimensionality of the moduli spaces ($Q N = Q' N'$).
The integer constant $N_0$ depends on the twist. This transformation 
provides an interesting tool  for studying (anti-)self-dual gauge fields on the torus.

In previous sections  we have expressed  certain self-dual potentials as an expansion
in the parameter $\Delta$.   It is henceforth interesting to analyse the
interplay of this
result with the Nahm transform. First of all we should find out the general
properties of the Nahm transform for the configurations in question.
 In our case the group is $SU(2)$ and the
configuration has nontrivial twist tensor $n_{0 3}= n_{1 2}=1$.  This
implies that the parameter $N_0=4$. Furthermore, the topological charge $Q$ of
these configurations is determined by the  twist matrices and
equals $Q=\frac{1}{2}$, as for the corresponding  constant field strength 
configuration.
Hence, according to the formulas given above, the Nahm dual is again an $SU(2)$ 
solution with topological charge   $Q'=\frac{1}{2}$. Now we can make use of the results
of Ref.~\cite{twist} to determine the twist tensor and torus size of the 
Nahm transformed field. Indeed, the Nahm dual twist tensor is equal to 
the original one, and the torus size is given by: $\frac{1}{2 l_0} \times
 \frac{1}{2 l_1} \times  \frac{1}{2 l_2} \times \frac{1}{2 l_3}$. Thus,
except for the different torus size,
 the Nahm transformed field is of the same type as   the original one. 
Furthermore, the size parameter $\Delta'$  of the Nahm transformed 
field is given by:
\be
\Delta'= -\Delta \quad .
\ee 
Therefore, the Nahm transform provides a  nonlinear relation for  our 
perturbative expansion. A full analysis of this point is difficult and
lengthy and will be left out from this paper, however it is instructive to 
look at the first few terms of this connection.

In order to construct the Nahm transform one has to study the  zero modes
of the Weyl equation in the fundamental representation of the group:
\be
\label{Weyleq}
(\overline{\cal D} -2 \pi \imath \bar{z})\,  \chi(x;z) = 0\quad ,
\ee
where $z_{\mu}$ represent the coordinates of a point in the Nahm dual torus.
The Weyl operator $\overline{\cal D}$ contains the self-dual potential
Eq.~(\ref{Adef}) which can be expanded in powers of $\sqrt{\lambda \Delta}$.
Although, for $\lambda \ne 1$ the configuration is not self-dual, it is still 
possible to define a Nahm transform (which will not be self-dual).
Thus, we can expand 
$\chi(x;z)$ in the same way and equate  to zero all of the powers of the 
equation separately. From  Eqs.~(\ref{Wexp}),(\ref{Sexp}) it is easy to see that 
the upper and lower components in colour space only mix for odd-even or even-odd powers
of the expansion parameter. Thus, we might write:
\be
\chi(x;z) = \left( \begin{array}{c} \phi_{+} + \sqrt{\lambda \Delta}
\phi'_{+} \\  \phi_{-} + \sqrt{\lambda \Delta} \phi'_{-}  \end{array}
\right)\quad .
\ee
In the previous formula, the explicit vector is in colour space, while
the quantities $\phi_{\pm}(x;z)$, $\phi'_{\pm}(x;z)$ are bi-spinors, which can be expanded in power series in $\lambda \Delta$.
Eq.~(\ref{Weyleq}) amounts for $\phi_{+}$ and $\phi'_{-}$ to the equations:
\bea
&(\overline{D}_{\half} -\imath S^{\dagger} - 2 \pi \imath \bar{z})\phi_{+} =\imath   \sqrt{\lambda \Delta} W_c^{\dagger} \phi'_{-} \\
&(\overline{D}_{- \half} +\imath S^{\dagger} - 2 \pi \imath \bar{z})\phi'_{-} =\imath   \frac{1}{\sqrt{\lambda \Delta}} W^{\dagger} \phi_{+}
\eea
and a similar equation holds for the remaining components. The symbol
$\overline{D}_{\half}$ is defined in (\ref{DQdef}). We see that in this 
way we get two independent solutions of (\ref{Weyleq}) as predicted by the 
index theorem. The Nahm transformed  $SU(2)$ vector potential is then given
by the formula:
\be
\label{NTdef}
\widehat{A}_{\mu}^{i j}(z) = \imath \int d^4 x \  \chi^{i \dagger} (x;z)
\frac{\partial}{\partial z_{\mu}} \chi^{j}(x;z)\quad , 
\ee
where the indices $i$,$j \in \{ 1,2\}$ label the two linearly independent and
orthonormal solutions. 

For the whole construction, the question of the boundary conditions satisfied 
by the spinors is crucial. Indeed, the naive  periodicity requirement:
\be
\phi_{\pm}(x+e_{\mu};z) = \exp \{ \pm \imath \frac{\pi}{2} \nmunu \frac{x_{\nu}}{l_{\nu}} \}\  \phi_{\pm}(x;z) 
\ee
is inconsistent. How to remedy this situation is what is studied in
Ref.~\cite{twist}. In the case at hand the easiest way out is to impose the periodicity 
requirement only for the $x_0$ and $x_1$ direction, while requiring only 
double period conditions on the other two. In short, this is just replicating 
the torus  in the  $x_3$ and $x_2$ directions. Consistently the integration in 
(\ref{NTdef}) has to be performed in this larger torus. 

To illustrate the procedure we will explicitly work out the lowest order 
term $\phi_{+}^{(0)}$, which satisfies:
\be
(\overline{D}_{\half}  - 2 \pi \imath \bar{z})\, \phi_{+}^{(0)}= 0   
\ee
This is a modification of the general equation studied in the previous
chapter for $q=\half$. Hence, following the same steps as before 
and imposing the new
boundary conditions, we arrive at a unique solution (up to  a 
multiplicative constant):
\be
\label{nahmzero}
\phi_{+}^{(0)} (x;z) = \exp\{ \pi \imath z_{\mu} y_{\mu})\}\ \widehat{\Psi}(y)
\ \left( \begin{array}{c} K'^{(0)} \\ 0 \end{array} \right)
\ee
where we have defined the auxiliary variable:
\be
y_{\mu} = x_{\mu} +2 l_{\mu} l_{\nu} n_{\mu \nu} z_{\nu}\quad ,
\ee
and the function $\widehat{\Psi}$ is the same one that appears in expression~(\ref{psiexp}),
with the replacement of $l_{2,3}$ by $2l_{2,3}$ and $\tau_{0,1}$ by $2\tau_{0,1}$.
The constant $K'^{(0)}$ is fixed by the normalisation condition. 
In the same way one derives for  $\phi_-^{(0)}$ the expression:
\be
\phi_{-}^{(0)} (x;z) = i\tau_2 \phi_+^{(0) \, *}(x;-z) \quad .
\ee

Now we might compute $\widehat{A}^{1 1}_{\mu}(x)$ by replacing
Eq.~(\ref{nahmzero}) in (\ref{NTdef}). Now introducing the complex variables:
\be
v_{\mu}= \frac{1}{l_{\mu}}(y_{\mu}+\imath n_{\mu \nu} y_{\nu}) 
\ee
we can express all derivatives with respect to $z_{\mu}$ in terms of
derivatives with respect to $v_0$, $v_1$, $v_0^*$ and  $v_1^*$.
For example:
\be
\label{PARTIALZ}
\frac{\partial}{\partial z_0} \phi_{+}^{(0)} = \left(\pi\imath y_0 -2\imath l_3\left(
\frac{\partial}{\partial v_0} - \frac{\partial}{\partial v_0^*}\right)\right)\phi_+^{(0)} \ .
\ee
 $\widehat{\Psi}(y)$ has a very simple dependence on $v^*_0$ 
 and hence one has:
 \be
\label{partvz}
 \frac{\partial}{\partial v^*_0}\phi_{+}^{(0)} = l_0\left(-\frac{\pi}{4 l_3}v_0
+\frac{\imath\pi}{2}\left(z_0+\imath z_3\right)\right) \phi_+^{(0)} \ .
\ee
The result of differentiating with respect to $v_0$ is much more complicated,
involving derivatives of Riemann's theta function. However, in the expression 
for $\widehat{A}$ it is possible to integrate by parts and make the 
derivatives with respect to $v_0$ act onto the complex conjugate of $ \phi_{+}^{(0)}$, 
for which the complex conjugate of Eq.~(\ref{partvz}) allows us to obtain a simple
expression. We end up with:
\bea
\nonumber
(\widehat{A}_{0}^{(0)})^{1 1}(z) &=& \imath \int d^4 x \   \phi_{+}^{(0) \dagger} 
\left(\pi \imath  y_{0} +4 \imath l_3 l_0 \Re\left\{-\frac{\pi }{4 l_3}  v_0 +
\frac{\imath\pi}{2} (z_0+\imath z_3)\right\}\right)\,   \phi_{+}^{(0)} \\
\label{NTlowest}
&=& 2 \pi l_0 l_3 z_3 \ .
\eea
In the right hand side of the first equality we have displayed  the contribution of the two
terms entering the right hand side of (\ref{PARTIALZ}). One can compute the
other components of the Nahm transformed field in the same fashion arriving
at:
\be
\widehat{A}_{\mu}^{(0)}(z)= -B'_{\mu}(z) 
\ee
where $B'_{\mu}(z)$ is given by the same expression~(\ref{Bdef}) as in section 2,
but with the lengths of the torus $l_{\mu}$, replaced by those of the Nahm
dual one $l'_{\mu}=\frac{1}{2 l_{\mu}}$. The previous result implies that
the Nahm dual of constant field strength configuration is a constant field
strength configuration, even if they are not self-dual, generalising the
result of   Ref.~\cite{vanbaal3}.

\section{Conclusions}

In the previous sections we have presented a systematic
expansion which allows the construction of self-dual $SU(2)$
Yang-Mills solutions with twist tensor $n_{0 3}=n_{1 2}=1$ on 
the torus. The size of higher order corrections depends on the 
lengths of the torus. For certain torus sizes the solution becomes
equal to the the well-known constant field-strength ones. The magnitude of
higher order corrections grows as we move away  from these torus sizes. 
We have also compared the landscape of the solution, as obtained from 
our analytical expressions to leading non-trivial order in the expansion,
to the numerical result obtained through standard techniques. The result 
is quite satisfactory at a qualitative and quantitative level. Finally, 
the expansion is used to obtain the Nahm transform of the self-dual
configuration. Curiously, the perturbative construction of the Nahm transform
has the same structure as the direct perturbative construction of the
self-dual configuration on the Nahm-self-dual torus. We have not been able to
equate these expansions order by order, but have shown this to be the case 
to lowest order. 

Let us now comment about the usefulness of our programme. Up to a proof, 
which we do not give, of convergence of our expansion, our method gives
a direct proof of the existence of the solutions and for them to have the
correct number of degrees of freedom. Even if calculating higher orders 
of the expansion turns out to be a difficult task, the expansion can be
of theoretical interest for different reasons. For example, it might allow
to investigate some general properties of the solutions. The case of the 
interplay with the Nahm transform is interesting, and should be pursued.
Furthermore, one can study certain extreme limits of the torus sizes, 
which might allow to obtain exact solutions.

Finally, we comment that our choice of $SU(2)$ and of the aforementioned
twist tensor has been dictated by simplicity. There is, however, no
a priori essential difficulty in generalising the construction given here
to other $SU(N)$ groups and different twist tensors.

\section*{Acknowledgements}
This work was financed by the CICYT under grant AEN97-1678.
We thank the Cen\-tro de Com\-pu\-ta\-ci\'on Cien\-t\'{\i}\-fi\-ca (UAM) for
the use of computing resources.

\end{document}